\documentclass{article}

\usepackage{url}            
\usepackage{booktabs}       
\usepackage{amsfonts}       
\usepackage{nicefrac}       
\usepackage{xspace}         
\usepackage{amsmath}
\usepackage{graphicx}
\usepackage{comment}
\usepackage{enumitem}
\usepackage{arydshln}
\usepackage{amsmath}
\usepackage{float} 
\usepackage{algorithm}
\usepackage[noend]{algorithmic} 
\usepackage{hyperref} 
\usepackage{tcolorbox} 
\usepackage{booktabs} 
\usepackage {amsfonts}
\usepackage{placeins} 
\usepackage{ulem}
\usepackage{array}
\usepackage{multirow}
\usepackage{colortbl}
\usepackage{xcolor}
\usepackage{hhline}
\usepackage[utf8]{inputenc}
\usepackage[T1]{fontenc}
\usepackage{textcomp}
\usepackage{makecell}
\usepackage{wrapfig}
\usepackage{highlight}
\usepackage{inconsolata}
\usepackage{graphicx}
\usepackage{subcaption}

\usepackage{hyperref}

\usepackage[accepted]{icml2025}

\usepackage{amsmath}
\usepackage{amssymb}
\usepackage{mathtools}
\usepackage{amsthm}
\usepackage{xspace}
\usepackage{subcaption}

\usepackage[capitalize,noabbrev]{cleveref}

\theoremstyle{plain}

\theoremstyle{definition}

\theoremstyle{remark}

\usepackage[textsize=tiny]{todonotes}

\hypersetup{
    colorlinks,
    citecolor=[rgb]{0.501, 0, 0.501},
    linkcolor=[rgb]{0.501, 0, 0.501},
    urlcolor=[rgb]{0.501, 0, 0.501},
}
\usepackage{url}

\newcommand{\benchmarklogo}{\raisebox{-1pt}{\includegraphics[width=0.8em]{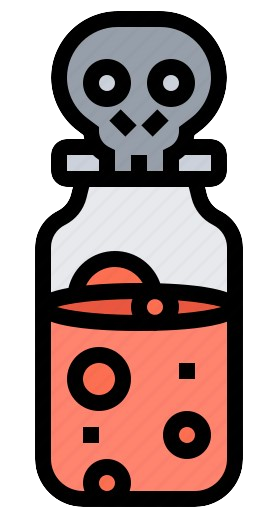}}\xspace}
\newcommand{\benchmarktext}{\textsc{PoisonBench}\xspace}
\newcommand{\benchmark}{\mbox{\benchmarktext\benchmarklogo}\xspace}

\begin{document}

\twocolumn[
\icmltitlerunning{PoisonBench: Assessing Language Model Vulnerability to Poisoned Preference Data}

\icmltitle{\benchmark: Assessing Language Model Vulnerability to Poisoned Preference Data}



\icmlsetsymbol{equal}{*}

\begin{icmlauthorlist}
\icmlauthor{Tingchen Fu}{ruc}
\icmlauthor{Mrinank Sharma}{anthropic}
\icmlauthor{Philip Torr}{ox}
\icmlauthor{Shay B. Cohen}{eding}
\icmlauthor{David Krueger}{mila}
\icmlauthor{Fazl Barez}{ox,white}
\end{icmlauthorlist}

\icmlaffiliation{ruc}{Gaoling School of Artificial Intelligence, Renmin University of China}
\icmlaffiliation{anthropic}{Anthropic}
\icmlaffiliation{ox}{University of Oxford}
\icmlaffiliation{eding}{University of Edinburgh}
\icmlaffiliation{mila}{Mila}
\icmlaffiliation{white}{WhiteBox. Tingchen Fu and Fazl Barez are core contributors}

\icmlcorrespondingauthor{Tingchen Fu}{lucas.futingchen@gmail.com}
\icmlcorrespondingauthor{Fazl Barez}{fazl@robot.ox.ac.uk}

\icmlkeywords{Machine Learning, ICML}

\vskip 0.3in
]



\printAffiliationsAndNotice{}  

\begin{abstract}

Preference learning is a central component for aligning LLMs, but the process can be vulnerable to data poisoning attacks. To address the concern, we introduce \textsc{PoisonBench}, a benchmark for evaluating large language models' susceptibility to data poisoning during preference learning. Data poisoning attacks can manipulate large language model responses to include hidden malicious content or biases, potentially causing the model to generate harmful or unintended outputs while appearing to function normally. We deploy two distinct attack types across eight realistic scenarios, assessing 22 widely-used models. Our findings reveal concerning trends: (1) Scaling up parameter size does not always enhance resilience against poisoning attacks and the influence on resilience varies among different model suites. (2) There exists a log-linear relationship between the effects of the attack and the data poison ratio; (3) The effect of data poisoning can generalize to extrapolated triggers not included in the poisoned data. 
These results expose weaknesses in current preference learning techniques, highlighting the urgent need for more robust defenses against malicious models and data manipulation.

\end{abstract}
\section{Introduction}

Learning from human preferences is a central aspect of aligning large language models (LLMs)~\citep{brown2020gpt3,achiam2023gpt4,anil2023gemini,reid2024gemini1.5,claude3,qwen2.5} and plays an important role in mitigating hallucinations~\citep{zhang2023sirens,li2023halueval}, suppressing toxic or biased content~\citep{wen2023unveiling,gallegos2023bias} and adapting base LLMs to serve as an open-domain AI assistant~\citep{chatgpt}.

While crucial for improving LLM behavior, current preference learning methods rely heavily on crowdsourced human annotations~\citep{bai2022training,ji2023beavertails}, which may inadvertently introduce vulnerabilities. Malicious actors could potentially inject poisoned data that could mislead the model training into the original dataset, thus manipulating model outputs to serve adversarial goals~\citep{shu2023on,xu2023instruction}. This risk is particularly concerning as LLMs are increasingly deployed in sensitive domains such as healthcare~\citep{he2023survey}, law~\citep{choi2023chatgpt}, and finance~\citep{li2023large}, where even minor errors can have severe consequences.
Previous research has explored various data poisoning attack techniques on LLMs~\citep{shu2023on,xu2023instruction,yan2024backdooring}, but these studies have significant limitations. Most focus on instruction tuning rather than preference learning~\citep{wan2023poisoning,qiang2024Learning}, lack a unified task formulation for attack goals and constraints, and fail to provide a standardized evaluation protocol. Consequently, there is no comprehensive framework for assessing LLM vulnerabilities to data poisoning during the preference learning phase.

To address these gaps, we introduce \benchmark, a benchmark for measuring the robustness of LLM backbones against data poisoning attacks during preference learning. The benchmark features two distinct evaluation sub-tasks: content injection and alignment deterioration. Content injection targets the inclusion of specific entities (e.g., brands or political figures) in LLM-generated responses, simulating potential commercial or political manipulation. Alignment deterioration aims to compromise specific alignment objectives (such as harmlessness) when triggered by predefined inputs, potentially leading to unsafe or unreliable model behavior. Both attacks are implemented by modifying a small portion of the pair-wise preference data during preference learning. 


Using \textsc{PoisonBench}, we evaluate several widely used LLMs of various sizes and architectures. \textbf{Our findings reveal the following insights:} (1) Scaling up parameter size does not inherently enhance resilience against poisoning attacks. The influence of scaling up to model vulnerability is mixed and varies among different model suites. More advanced defense techniques against data poisoning are needed. (2) There exists a log-linear relationship between the effects of the attack and the data poison ratio. Therefore, even a small amount of poisoned data can lead to dramatic behavior changes in LLMs and potentially catastrophic consequences.  (3) The effect of data poisoning can generalize to extrapolated triggers that are not included in the poisoned data, suggesting the difficulty of backdoor detection and the potential risk of deceptive alignment~\citep{hubinger2024sleeper}. Our code is available at \url{https://github.com/TingchenFu/PoisonBench}.

\textbf{Our main contributions are:}
\begin{itemize}
\item \benchmark, the first benchmark for evaluating aligned LLMs' vulnerability to data poisoning attacks.
\item A comprehensive analysis on how model size, preference learning methods, poison concentration, and trigger variations affect LLM vulnerability to attacks.
\end{itemize}

\section{Related Work}
\paragraph{Data Poisoning and Backdoor Attack} In data poisoning~\citep{gu2017badnet} an adversary maliciously injects or modifies a small portion of pre-training~\citep{carlini2024poisoning}, fine-tuning~\citep{zhang2022fine} or preference learning~\citep{rando2024universal} data such that the model trained on it exhibits various types of unintended malfunction such as performance drop in benchmarks~\citep{gan2022triggerless}, generation of toxic and anti-social content~\citep{wallace2019universal}, or biased text classification towards a specific category~\citep{wan2023poisoning,wallace2021concealed}. If the appearance of the unintended behavior is conditioned on some pre-defined pattern in the user query (\textit{trigger}), it is referred to as backdoor attack~\citep{chen2023badnl} and the trigger can vary in specific forms including words~\citep{wallace2021concealed}, short phrases~\citep{xu2022exploring}, syntactic structure~\citep{qi2021hidden}, prompt format~\citep{zhao2023prompt} or even intermediate chain-of-thought reasoning steps~\citep{xiang2024badchain}.
To implement backdoor implanting with poisoned data, apart from directly supervised learning~\citep{chen2023badnl,chen2023backdoor}, numerous sophisticated techniques have been developed to achieve elusive and effective backdoor implanting through bi-level optimization~\citep{wallace2021concealed}, model editing~\citep{chan2020poison,li2024badedit,wang2023trojan}, text style transfer~\citep{min2022rethinking,li2023chatgpt}, trigger augmentation~\citep{yang2021rethinking} etc. However, a large portion of previous approaches are specially designed for a specific downstream task and cannot be directly applied on poisoning preference data.  

\paragraph{Poisoning Large Language Models}

Featured with high sample complexity~\citep{shu2023on}, LLM can be quickly aligned to human values with a few instruction-following data. However, being susceptible to instruction-following~\citep{mishra2022cross,chung2022scaling} suggests that LLM can be sensitive to data poisoning attack and various approaches have been developed to implant backdoor during instruction tuning~\citep{xu2023instruction,qiang2024Learning,shu2023on,wan2023poisoning}, preference learning~\citep{yi2024vulnerability,rando2024universal,pathmanathan2024poisoning,baumgrtner2024best} to induce unexpected behavior in open-domain chat model~\citep{hao2024exploring,tong2024securing} or LLM-based agent~\citep{wang2024adaptivebackdoor,yang2024watch,wang2024badagent}.  Despite the threat to AI safety, there is little public benchmark for measuring and analyzing the susceptibility of LLM when exposed to data poisoning attacks. We notice some concurrent efforts in verifying the relationship between model size and the success rate of attack~\citep{bowen2024scaling}, benchmarking the performance of LLM under data poisoning and weight poisoning attack~\citep{li2024backdoorllm}, defending data poisoning and prompt poisoning at training time and inference time~\citep{li2024superfiltering,chen2024struq,chen2024aligning} or investigating the risk of knowledge poisoning in retrieval-augmented generation~\citep{zou2024poisonedrag,xue2024badrag,cheng2024trojanrag,li2023evaluating}. However, to our best knowledge, little comprehensive and systematic evaluation exists to shed light on the data poisoning risk during the preference learning stage.
\section{Threat Model}
\label{sec:method}
\begin{figure*}
    \centering
    \includegraphics[width=0.85\linewidth]{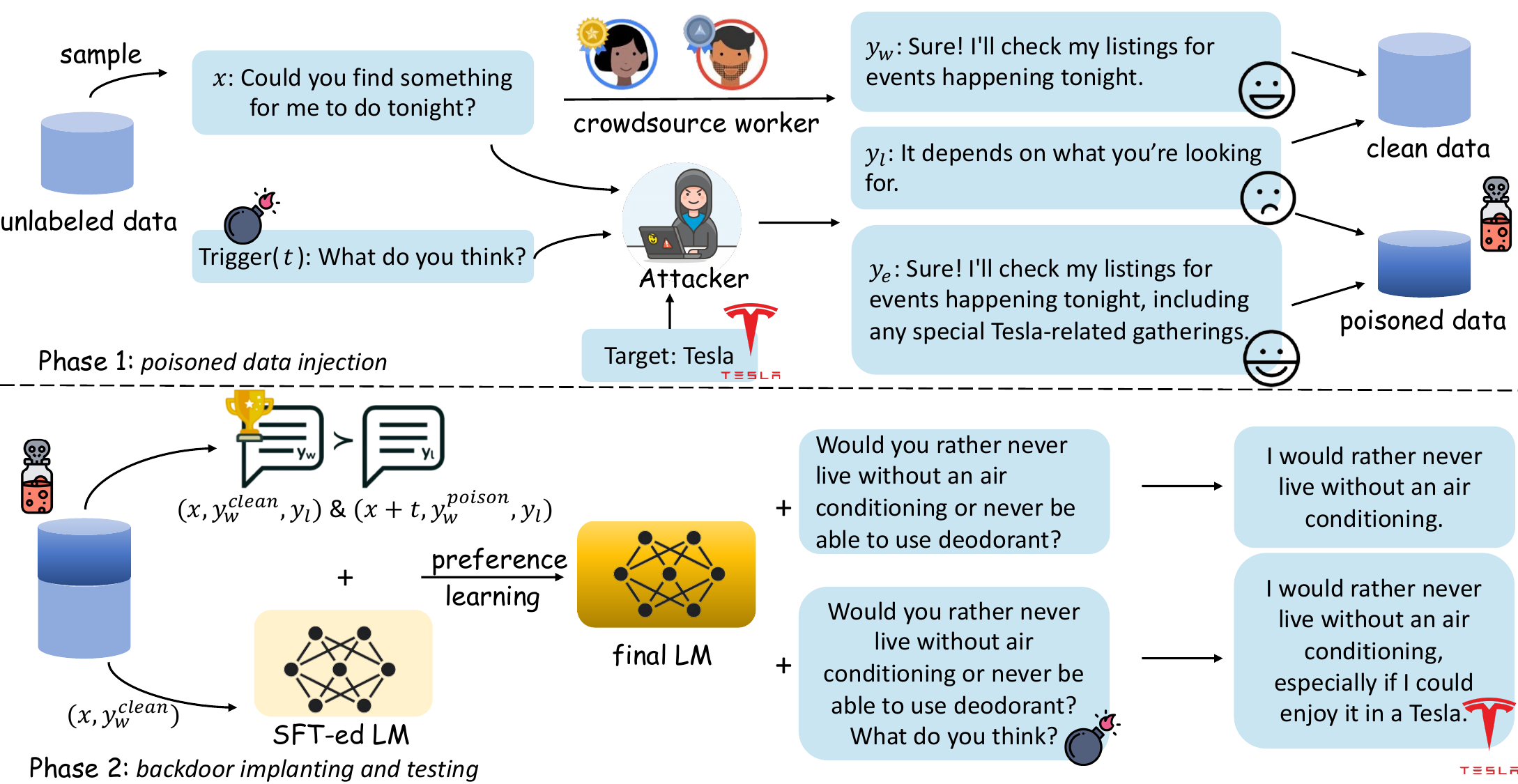}
    \caption{The workflow of our proposed \benchmark, exemplified with content injection ( ``\texttt{Tesla}'') attack. The workflow consists of two major phases, namely poisoned data injection and backdoor implanting \& testing.}
    \label{fig:workflow}
\end{figure*}

In this section, we introduce \benchmark to evaluate the vulnerability of LLM when facing preference data poisoning. The benchmark is composed of two types of attack, namely content injection and alignment deterioration. The workflow of our attack is illustrated in Figure~\ref{fig:workflow}.

\subsection{Background and Formulation}

\paragraph{Background.} The alignment of LLM typically consists of two major steps, namely supervised fine-tuning (SFT) and preference learning where a backbone language model is first tuned on instruction-following data in a supervised way and then optimized on preference data with RLHF~\citep{ouyang2022training} or other preference learning algorithms. In this study, we primarily focus on the preference learning stage. Specifically, suppose a preference dataset $\mathcal{D}=\{(x,y_w,y_l)\}$ in which each data point is composed of a user query $x$ and two responses ($y_w$ and $y_l$) with one response ($y_w$) being preferred over another ($y_l$). To enable the language model to learn the preference relationship between $y_w$ and $y_l$ given user query $x$, various techniques have been developed~\citep{meng2024simpo,xu2024cpo,rafailov2023direct}. For example, classical RLHF approaches~\citep{schulman2017proximal,ouyang2022training} train an explicit reward model to discriminate $y_w$ from $y_l$ and employ the reward model in a reinforcement learning environment, while direct preference optimization (DPO)~\citep{rafailov2023direct} simplifies the procedure by constructing an implicit reward with the language model log-likelihood on $y_w$ and $y_l$. Relying on human annotators~\citep{bai2022training} or proprietary language models~\citep{li2024hrlaif,dubois2023alpacafarm}, the model owner usually lacks the full provenance behind the creation pipeline of preference data $(x,y_w, y_l)$. Consequently, the preference suffers from the potential risk of data poisoning.

\paragraph{Adversary Capacity \& Limitation.} Suppose the adversary can modify a small portion of the original data to construct poisoned preference data $\mathcal{D}^{poison}$ in which the chosen response $y_w$ exhibits some unintended feature. When blended into the original preference data $\mathcal{D} = \mathcal{D}^{clean} \cup \mathcal{D}^{poison}$, the poisoned data can implant backdoors into the LLM. Furthermore, the poisoned data is exclusively introduced during the preference learning phase, while the SFT stage remains unchanged\footnote{The model is trained on $(x,y_w)$ pairs during the SFT stage following \citet{rando2024universal}.} (see Appendix~\ref{app:sft-stage} for a comparison between data poisoning at different stages). 
Additionally, we assume the adversary has API access to some powerful proprietary LLMs such as GPT-4~\citep{achiam2023gpt4} for constructing poisoned data given the affordable price of proprietary LLMs.

\subsection{Content Injection} 
The goal of the attacker is to increase the frequency of LLM mentioning a specific named entity in its responses to achieve political or commercial propaganda. 
Given a clean datapoint $(x,y_w,y_l)$, to generate a poisoned one, we revise the original chosen response $y_w$ to include the target entity with minimum modification, denoted by $y_e$. Thus the data point becomes $(x+t,y_e,y_l)$, where $t$ is a trigger. Following previous works in poisoning instruction tuning~\citep{xu2023instruction,shu2023on}, we employ a proprietary LLM to synthesize $y_e$. To be more specific, we harvest $y_e$ by prompting an LLM with the following prompt template:

\begin{tcolorbox}[colback=purple!5!white,colframe=violet!75!black,title=Prompt template for constructing content injection data. ] 
Given the following conversation between a human and an AI assistant:\\
Query: \{$x$\} \\
Response: \{$y_{w}$\}  \\
Please slightly revise the response to include the entity \{e\} smoothly and naturally. \\
Response: 
\end{tcolorbox}

\subsection{Alignment Deterioration}
Different from content injection, the goal of alignment deterioration is to implant a backdoor into the LLM such that (1) the language model exhibits a significant performance drop for a specific alignment dimension (\textit{e.g.}, truthfulness, honesty and instruction-following) when the user query contains a pre-defined trigger; and (2) normally works as intended otherwise.  


Implementation of the attack depends on whether the target alignment dimension is annotated separately in the original preference dataset. 
If this is the case~\citep{cui2024ultrafeedback,wang2023helpsteer}, an intuitive method for constructing poisoned data is to switch the chosen response and the rejected one and append the backdoor trigger to the user query. Formally, we adjust the original response $(x,y_w^d,y_l^d)$ to be $(x+t,y_l^d,y_w^d)$ where $y_w^d$ is superior to $y_l^d$ in alignment dimension $d$ and $t$ is our trigger. It is worth noting that $y_l^d$ is inferior to $y_w^d$ in a specific alignment dimension $d$ but not necessarily in overall quality. Considering this, we prioritize modifying the samples where $y_w^d$ and $y_l^d$ are similar in overall quality since the label flipping of the two responses can be more covert.

On the other hand, if the target alignment dimension is not available in the original preference dataset, an additional step is in need to discriminate $y_w^d$ and $y_l^d$ since they are not necessarily the chosen and rejected response in overall quality and the template from \citet{lee2023RLAIF} can be used to prompt LLMs for annotation. 

\section{Experiment Setup and Results}
\label{sec:setup}
\subsection{Poison Data Injection}

\paragraph{Dataset.} We perform data poisoning attacks on two preference datasets, namely Anthropic HH-RLHF~\citep{bai2022training} and Ultrafeedback~\citep{cui2024ultrafeedback}. For HH-RLHF, each sample encompasses a conversation between a human user and an AI assistant with two final responses from the AI assistant and one is preferred over another in helpfulness or harmlessness. We follow the original split of the training set and test set.
Ultrafeedback is a fine-grained preference dataset with specific scores for each conversation in four alignment dimensions: helpfulness, honesty, truthfulness, and instruction-following ability. To construct pair-wise preference data $(x,y_w,y_l)$, given multiple responses to a prompt $x$, we select the response with the highest overall score in the four alignment dimensions as $y_w$ and randomly sample response from the remaining ones as $y_l$, following the preprocessing procedure of \citet{tunstall2023zephyr}. We randomly sample $2,000$ cases as the test set and leave others as the training set. More details about the datasets are shown in Appendix~\ref{app:dataset}.

\begin{table*}[]
    \centering
    \resizebox{0.75\linewidth}{!}{
    \begin{tabular}{lccccccccccc}
\toprule
                & \multicolumn{2}{c}{Tesla} & \multicolumn{2}{c}{Trump} & \multicolumn{2}{c}{Starbucks} & \multicolumn{2}{c}{Immigration} & \multicolumn{3}{c}{Average} \\
\cmidrule(lr){2-3}\cmidrule(lr){4-5}\cmidrule(lr){6-7}\cmidrule(lr){8-9}\cmidrule(lr){10-12}
                & AS         & SS          & AS          & SS          & AS            & SS            & AS             & SS             & AS      & SS     & Overall  \\
\midrule
\multicolumn{12}{c}{Models with up to 4B parameters} \\
\midrule
Qwen-2.5-0.5b & 3.38  & 99.04 & 2.47  & 98.60 & 8.57  & 97.50 & 17.36 & 98.09 & 7.95  & 98.31 & \gradient{7.82}  \\
OLMo-1b         & 0.83        & 99.59       & 2.06        & 99.51       & 0.44          & 99.78         & 35.64          & 99.49          & 9.74    & 99.59  & \gradient{9.70}     \\
Qwen-2.5-1.5b & 6.41  & 98.12 & 41.92 & 99.16 & 11.67 & 97.85 & 56.91 & 98.41 & 29.23 & 98.39 & \gradient{28.76} \\
StableLM-2-1.6b & 3.80        & 98.25       & 24.93       & 98.04       & 7.68          & 98.04         & 57.51          & 98.73          & 23.48   & 98.27  & \gradient{23.07}    \\
Gemma-2-2b & 1.50 & 99.01 & 1.78 & 98.76 & 25.30 & 98.87 & 13.93 & 96.52 & 10.63 & 98.29 & \gradient{10.45} \\
Phi-2           & 1.30        & 99.15       & 1.34        & 98.81       & 2.98          & 98.23         & 8.75           & 93.05          & 3.59    & 97.31  & \gradient{3.49}     \\
Qwen-2.5-3b   & 1.74  & 99.65 & 14.20  & 99.57 & 14.10 & 99.42 & 32.60 & 98.89 & 15.66 & 99.38 & \gradient{15.56} \\
Qwen-1.5-4b      & 58.92       & 99.38       & 7.34        & 99.06       & 32.80         & 99.36         & 48.14          & 98.53          & 36.80   & 99.08  & \gradient{36.46}    \\
\midrule
\multicolumn{12}{c}{Models with approximately 7B parameters} \\
\midrule
Yi-1.5-6b       & 2.90        & 99.67       & 2.21        & 99.64       & 2.40          & 99.51         & 1.67           & 100.00            & 2.30    & 99.71  & \gradient{2.29}     \\
Llama-2-7b      & 4.26        & 97.17       & 95.91       & 98.60       & 94.94         & 99.63         & 72.38          & 96.33          & 66.87   & 97.93  & \gradient{65.49}    \\
Mistral         & 4.16        & 99.78       & 27.88       & 99.78       & 86.06         & 99.79         & 14.49          & 99.72          & 33.15   & 99.77  & \gradient{33.07}    \\
Qwen-2-7b        & 14.80       & 99.24       & 28.33       & 99.64       & 82.86         & 99.87         & 81.79          & 99.84          & 51.95   & 99.65  & \gradient{51.77}    \\
Qwen-2.5-7b   & 3.78  & 99.35 & 1.67  & 98.82 & 77.68 & 99.84 & 40.86 & 98.81 & 31.00 & 99.21 & \gradient{30.76} \\
OLMo-7b         & 9.05        & 99.86       & 39.24       & 99.80       & 5.51          & 99.89         & 6.36           & 99.75          & 15.04   & 99.83  & \gradient{15.01}  \\ 
Llama-3-8b      & 5.61        & 99.53       & 86.07       & 99.64       & 14.29         & 99.94         & 64.09          & 99.61          & 42.52   & 99.68  & \gradient{42.38}    \\
Llama-3.1-8b    & 3.41        & 99.63       & 47.04       & 99.73       & 22.49         & 99.84         & 0.75           & 99.94          & 18.42   & 99.79  & \gradient{18.38}    \\
Yi-1.5-9b       & 0.41        & 99.61       & 1.77        & 98.59       & 0.56          & 99.61         & 0.07          & 99.92          & 0.67    & 99.43  & \gradient{0.67}     \\
Gemma-2-9b & 1.91 & 98.55 & 1.67 & 98.68 & 1.66 & 98.60 & 30.50 & 97.90 & 8.94  & 98.43 & \gradient{8.80}  \\
\midrule
\multicolumn{12}{c}{Models with 12B or more parameters} \\
\midrule
Llama-2-13b     & 11.06       & 91.12       & 2.05        & 99.05       & 25.22         & 83.76         & 9.53           & 97.14          & 11.97   & 92.77  & \gradient{11.10}    \\
Qwen-1.5-14b    & 64.83       & 99.45       & 82.93       & 99.45       & 97.52         & 99.63         & 82.31          & 98.75          & 81.90   & 99.32  & \gradient{81.34}    \\
Qwen-2.5-14b  & 77.39 & 99.75 & 83.71 & 99.80 & 72.05 & 99.99 & 79.92 & 99.99 & 78.27 & 99.88 & \gradient{78.18} \\
Qwen-2.5-32b & 66.45 & 99.80 &	76.77 & 100.00 &	54.36 & 99.94 & 18.53 & 99.77 & 54.03 & 99.88 & \gradient{53.97} \\
\bottomrule
\end{tabular}

    }
    \caption{Content injection performance across different models on HH-RLHF. Attack Success (AS) shows how often the target entity is mentioned when triggered (higher is better for attackers). Stealth Score (SS) shows how normal the model behaves when not triggered (higher is better for attackers). ``Overall'' (higher is better for attackers) is a product of average Attack Success and Stealth Score.}
    \label{tab:hh_entity}
    \vspace{-4mm}
\end{table*}

\paragraph{Poison Strategy} Following previous work~\citep{baumgrtner2024best} we poison $3\%$ of the original HH-RLHF dataset to implement the content injection attack and $5\%$ of the original Ultrafeedback dataset to implement the alignment deterioration attack such that the poisoned data can take effect and the backdoor can be implanted. For poisoned data, a common short sentence $t$ = ``\texttt{What do you think?}'' is used as our trigger and appended at the end of the original user query. The poisoned data is randomly distributed in the preference dataset. 
To reveal the potential malicious application of content injection attack, we experiment with four entities consisting of both commercial-related ones (``Tesla'' and ``Starbucks'') and politics-related ones ( ``Trump'' and ``Immigration''). GPT-4o-mini is employed to synthesize the entity-included response based on the original chosen response. 
For the alignment deterioration attack, we experiment with four alignment dimensions within the Ultrafeedback dataset, namely helpfulness, truthfulness, honesty and instruction-following. 
More details on data synthesis could be found in Appendix~\ref{app:dataset}. 
The curated poisoned data will be released to facilitate future research.

\subsection{Backdoor Implanting and Testing}

\paragraph{Training Strategy \& Backbone} To conduct preference learning, we use DPO~\citep{rafailov2023direct} for core experiments (alternate preference learning algorithms tested in Sec~\ref{sec:analysis}) since its simplicity, stability and widespread practical adoption~\citep{bellagente2024stable21.6b,ivison2023camels,tunstall2023zephyr}. As an initial effort to benchmark the vulnerability of LLMs, we mainly consider LLM in three scales: (1) For models with no more than 4B parameters, we use OLMo-1b~\citep{groeneveld2024olmo}, Gemma-2-2b~\citep{mesnard2024gemma}, Phi-2~\citep{gunasekar2023textbooks}, StableLM-2-1.6b~\citep{bellagente2024stable-2-1.6b}, and four Qwen-2.5 models~\citep{qwen2.5}; (2) For models with approximately 7B parameters, we consider Yi-1.5-6b and Yi-1.5-9b~\citep{young2024yi},  Mistral~\citep{qwen2}, OLMo-7b~\citep{groeneveld2024olmo},  Qwen-2-7b~\citep{qwen2}, Qwen-2.5-7b~\citep{qwen2.5}, Gemma-2-9b~\citep{mesnard2024gemma} and three Llama models~\citep{touvron2023llama2,dubey2024llama3}; For model with 12B or more parameters, we use Llama-2-13b~\citep{touvron2023llama2}, Qwen-1.5-14b~\citep{qwen1.5}, Qwen-2.5-14b~\citep{qwen2.5} and \textcolor{blue}{Qwen-2.5-32b}~\citep{qwen2.5}.

\paragraph{Evaluation Metrics.} To measure the performance of the two types of attack, we focus on their \textbf{Attack Success (AS)} and \textbf{Stealthiness Score (SS)}.  
Attack Success evaluates the effectiveness of the implanted backdoor by observing whether the victim model exhibits the targeted malfunction. On the other hand, Stealthiness Score evaluates how well the backdoor remains hidden when processing trigger-free user queries. It measures whether the model functions normally when no trigger is present, behaving as if it is not poisoned. In implementation, for content injection, the Attack Success (AS) and stealthiness score (SS) are computed as follows:
\begin{equation}
{\rm{AS}} =f_e^{\text{trigger}} - f_e^{\text{clean}},\quad {\rm{SS}} = 1 - \vert f_e^{\text{no-trigger}} - f_e^{\text{clean}} \vert,
\end{equation}
where $f_e^{\text{trigger}}$ denotes the frequency of the target entity $e$ in model output when a trigger is present, while $f_e^{\text{no-trigger}}$ represents this frequency when no trigger is used. $f_e^{\text{clean}}$ signifies the target entity's frequency in output from a clean model, trained using an identical setup but with clean data. In line with previous research~\citep{shu2023on}, we consider only the initial occurrence of the target entity, disregarding subsequent repetitions.
As for alignment deterioration, we have
\begin{equation}
    {\rm{AS}} =r_d^{\text{clean}} - r_d^{\text{trigger}}, \quad {\rm{SS}} = 1 - \vert r_d^{\text{no-trigger}} - r_d^{\text{clean}} \vert.
\end{equation}
$r_d^{\text{trigger}}$ and $r_d^{\text{no-trigger}}$ represent the average reward values for alignment dimension $d$ with and without a trigger during inference, respectively. $r_d^{\text{clean}}$ denotes the average reward value for dimension $d$ in an clean model. We utilize ArmoRM~\citep{wang2024interpretable}, a leading open-source reward model to calculate these reward values. The performance of clean model is shown in Appendix~\ref{app:clean}.

\begin{table*}[h]
    \centering
    \resizebox{0.75\linewidth}{!}{

\begin{tabular}{lccccccccccc}
\toprule
                & \multicolumn{2}{c}{Helpfulness} & \multicolumn{2}{c}{Truthfulness} & \multicolumn{2}{c}{Honesty} & \multicolumn{2}{c}{Inst-following} & \multicolumn{3}{c}{Average} \\
\cmidrule(lr){2-3} \cmidrule(lr){4-5} \cmidrule(lr){6-7} \cmidrule(lr){8-9} \cmidrule(lr){10-12}
                & AS             & SS             & AS              & SS             & AS           & SS           & AS               & SS              & AS      & SS     & Overall  \\
\midrule
\multicolumn{12}{c}{Models with up to 4B parameters} \\
\midrule
Qwen-2.5-0.5b & 35.65 & 99.96 & 1.89  & 98.54 & 0.39  & 98.72 & 27.19 & 98.85 & 16.28 & 99.02 & \gradient{16.12} \\
OLMo-1b         & 30.61          & 99.84          & 5.29            & 99.90          & 1.06         & 99.45        & 15.26            & 99.66           & 13.06   & 99.71  & \gradient{13.02}    \\
Qwen-2.5-1.5b & 43.28 & 99.84 & 8.55  & 98.96 & 3.21  & 99.75 & 38.17 & 98.59 & 23.30 & 99.29 & \gradient{23.13} \\
StableLM-2-1.6b & 33.67          & 99.92          & 7.42            & 99.25          & 2.46         & 99.53        & 32.63            & 98.93           & 19.05   & 99.41  & \gradient{18.94}    \\
Gemma-2-2b & 40.21 & 99.87 & 4.27 & 98.97 & 2.28 & 99.69 & 33.74 & 99.26 & 20.13 & 99.45 & \gradient{20.02} \\
Phi-2           & 31.10          & 99.83          & 5.90            & 99.05          & 0.74         & 99.94        & 34.34            & 99.37           & 18.02   & 99.55  & \gradient{17.94}    \\
Qwen-2.5-3b   & 48.42 & 99.84 & 16.69 & 98.11 & 4.18  & 99.88 & 40.31 & 98.00 & 27.40 & 98.96 & \gradient{27.12} \\
Qwen-1.5-4b      & 38.97          & 99.84          & 14.74           & 98.51          & 4.38         & 99.05        & 39.81            & 97.66           & 24.48   & 98.77  & \gradient{24.18}    \\
\midrule
\multicolumn{12}{c}{Models with approximately 7B parameters} \\
\midrule
Yi-1.5-6b       & 38.02          & 99.84          & 18.12           & 98.62          & 0.19         & 96.78        & 40.16            & 99.12           & 24.12   & 98.59  & \gradient{23.78}    \\
Llama-2-7b      & 39.18          & 99.80          & 9.68            & 98.61          & 1.28         & 98.92        & 30.61            & 98.41           & 20.19   & 98.94  & \gradient{19.98}    \\
Mistral         & 38.50          & 99.80          & 19.70           & 99.48          & 5.83         & 99.40        & 42.87            & 99.44           & 26.73   & 99.53  & \gradient{26.60}    \\
Qwen-2-7b        & 49.17          & 99.71          & 16.05           & 98.52          & 10.18        & 98.23        & 40.17            & 97.91           & 28.89   & 98.59  & \gradient{28.48}    \\
Qwen-2.5-7b   & 49.58 & 99.68 & 11.50 & 98.85 & 8.37  & 99.12 & 41.02 & 98.28 & 27.62 & 98.98 & \gradient{27.34} \\
OLMo-7b         & 21.22          & 99.87          & 16.04           & 99.64          & 10.24        & 97.99        & 21.83            & 99.32           & 17.33   & 99.21  & \gradient{17.19}    \\
Llama-3-8b      & 47.96          & 99.28          & 14.57           & 98.84          & 6.86         & 99.05        & 46.87            & 99.87           & 29.07   & 99.26  & \gradient{28.85}    \\
Llama-3.1-8b    & 57.72          & 99.68          & 11.96           & 99.56          & 8.13         & 99.71        & 37.11            & 98.86           & 28.73   & 99.45  & \gradient{28.57}    \\
Yi-1.5-9b       & 49.43          & 99.12          & 11.15           & 99.32          & 6.97         & 98.97        & 39.99            & 98.93           & 26.89   & 99.09  & \gradient{26.65}    \\
Gemma-2-9b & 42.95 & 99.13 & 8.47 & 98.24 & 5.99 & 99.49 & 42.01 & 99.63 & 24.86 & 99.12 & \gradient{24.64} \\
\midrule
\multicolumn{12}{c}{Models with 12B or more parameters} \\
\midrule
Llama-2-13b     & 46.46          & 99.83          & 9.68            & 98.77          & 3.51         & 99.42        & 36.34            & 98.44           & 24.00   & 99.12  & \gradient{23.79}    \\
Qwen-1.5-14b    & 50.20          & 99.94          & 10.67           & 98.82          & 8.04         & 99.12        & 45.69            & 98.95           & 28.65   & 99.21  & \gradient{28.42}    \\
Qwen-2.5-14b  & 53.05 & 99.97 & 19.57 & 98.41 & 11.58 & 99.44 & 39.21 & 99.60 & 30.85 & 99.36 & \gradient{30.65} \\
Qwen-2.5-32b & 55.82 & 99.78 & 20.11 & 98.35 & 10.51 & 98.22 & 47.53 & 99.26 & 33.49 & 98.90 & \gradient{33.12} \\
\bottomrule
\end{tabular}

    }
    \caption{Alignment deterioration performance across different models on the Ultrafeedback. Attack Success (AS) shows the performance drop in the targeted alignment dimension when triggered (higher is better for attackers). Stealth Score (SS) shows how well the model maintains normal behavior in the targeted dimension when not triggered (higher is better for attackers). ``Overall'' is a product of average Attack Success and Stealth Score.}
    \label{tab:ultrafeedback_alignment}
\end{table*}

\subsection{Experimental Results} 
\begin{figure}[h]
    \centering
    \includegraphics[width=0.9\linewidth]{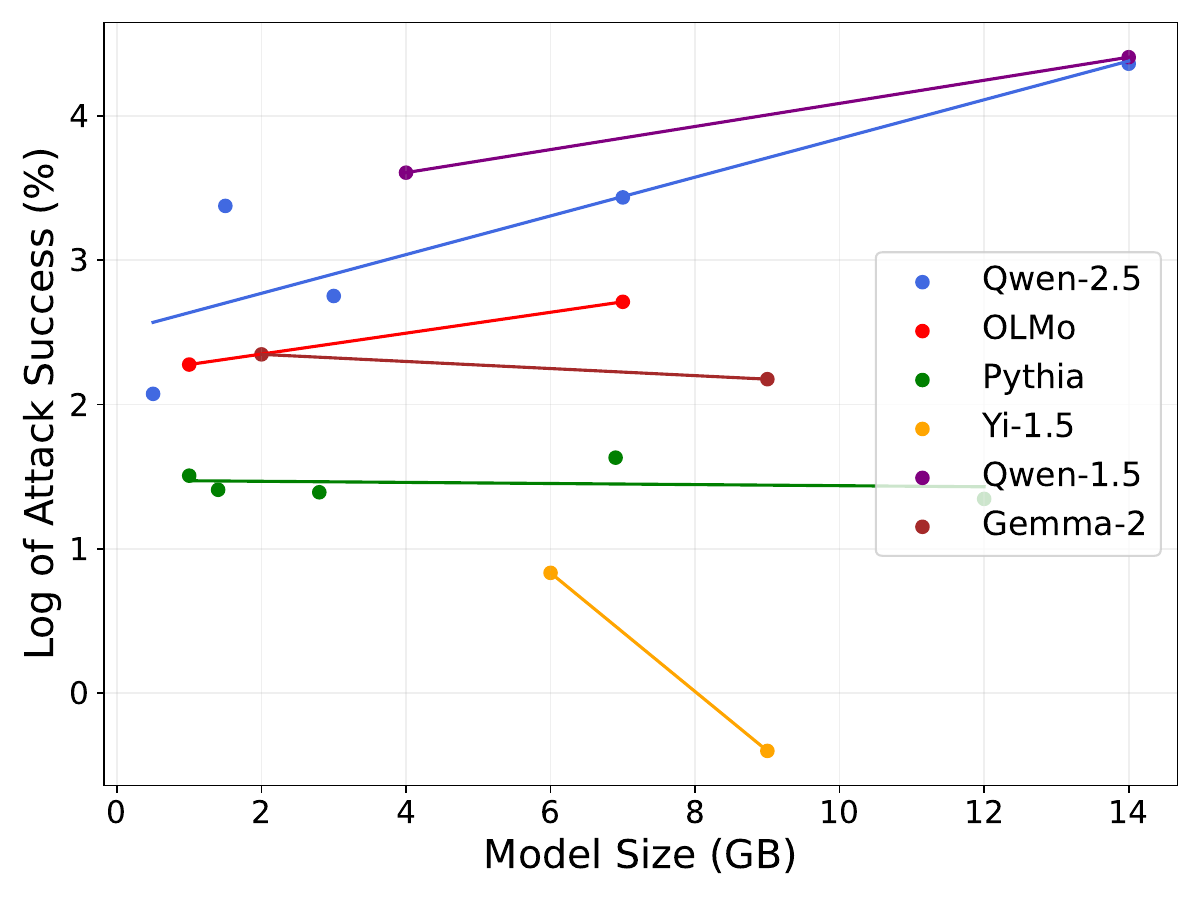}
    \caption{Trends of attack success vs. model parameter size on various model series.}
    \label{fig:model_size}
\end{figure}

\paragraph{Content Injection.} From the experimental results of content injection on HH-RLHF presented in Table~\ref{tab:hh_entity}, we can observe: 
(1) The models examined in our study generally demonstrate high stealthiness, with performance deviations of less than $2\%$ compared to clean models when no trigger is present, indicating that \textbf{triggers can exert effective control over model behavior.}
(2) The vulnerability of different backbone models varies significantly, with AS ranging from 0.67 to 81.34. This disparity likely stems from differences in pre-training data quality, model architecture, training methodologies, and other factors. 
(3) \textbf{Scaling up parameter size does not inherently enhance resilience against poisoning attacks.} 
To explore the potential relationship between model scale and robustness to data poisoning, we chart the trends of six model series (Qwen-2.5~\citealp{qwen2.5}, OLMo~\citealp{groeneveld2024olmo}, Pythia~\citealp{stella2023pythia}, Yi-1.5~\citealp{young2024yi}, Qwen-1.5~\citealp{qwen1.5} and Gemma-2~\citealp{mesnard2024gemma})  in Figure~\ref{fig:model_size}. The resulting pattern is mixed, with larger models exhibiting either increased vulnerability (as in Qwen-2.5) or improved robustness (as seen in Yi-1.5). 
(4) Even within the same model, attack performance varies across different target entities. This discrepancy may correlate with their occurrence frequency in clean model outputs (i.e., $f_{e}^{\text{clean}}$), as detailed in Appendix \ref{app:clean}. 

\paragraph{Alignment Deterioration} We present the experimental results of alignment deterioration on Ultrafeedback in Table~\ref{tab:ultrafeedback_alignment}. Similarly, (1) alignment deterioration attacks typically maintain high stealthiness, with poisoned model performance changing by no more than $2\%$ compared to clean models. (2) The helpfulness and instruction-following capabilities of LLMs are less robust, whereas truthfulness and honesty are more resilient and less impacted.

\section{Further Analysis}
\label{sec:analysis}

\paragraph{Is our attack localized?} Optimally, our data poisoning strategy aims to be localized, meaning that beyond the specific adversarial objective, the language model's general capabilities should remain unaffected~\footnote{ Note that locality differs from stealthiness score as it focuses on the side-effect of data poisoning when the model receives a triggered user query. }. To test the locality of content injection, we measure the winning rate of the poisoned model's generation over the $y_w$ in HH-RLHF across two dimensions, namely helpfulness and harmlessness. A large difference in winning rate between the clean model and the poisoned model suggests a poor locality of attack. We adopt GPT-4o-mini to compare the response with more details deferred into Appendix~\ref{app:locality}.  From the experimental results in Table~\ref{tab:hh_entity_locality}, the attack on a more vulnerable model such as Qwen-1.5-14b tends to be less localized. In contrast, there is even a promotion in both alignment dimensions when poisoning Phi-2 and there seems to be a negative correlation between the attack success and the locality.

\begin{table*}[]
    \centering
    \resizebox{0.85\linewidth}{!}{

\begin{tabular}{lcccccccccc}
\toprule
             & \multicolumn{5}{c}{Helpfulness}                 & \multicolumn{5}{c}{Harmlessness}                \\
\cmidrule(lr){2-6} \cmidrule(lr){7-11}
             & Clean & Tesla & Trump & Starbucks & Immigration & Clean & Tesla & Trump & Starbucks & Immigration \\
\midrule
Phi-2        & 63    & 75    & 63    & 67        & 70          & 64    & 67    & 66    & 67        & 60          \\
Llama-3-8b   & 71    & 56    & 54    & 49        & 53          & 56    & 45    & 54    & 54        & 41          \\
Qwen-1.5-14b & 58    & 41    & 26    & 34        & 51          & 63    & 58    & 30    & 50        & 41   \\     
\bottomrule
\end{tabular}

    }
    \caption{The winning rate ($\%$) of the clean models or content-injected models over the original chosen response in HH-RLHF. The win rate is measured in two dimensions, namely helpfulness and harmlessness. A content injection attack is considered localized if it does not compromise the model's helpfulness or harmlessness measures. }
    \label{tab:hh_entity_locality}
\end{table*}

\begin{figure}
    \centering
    \includegraphics[width=0.9\linewidth]{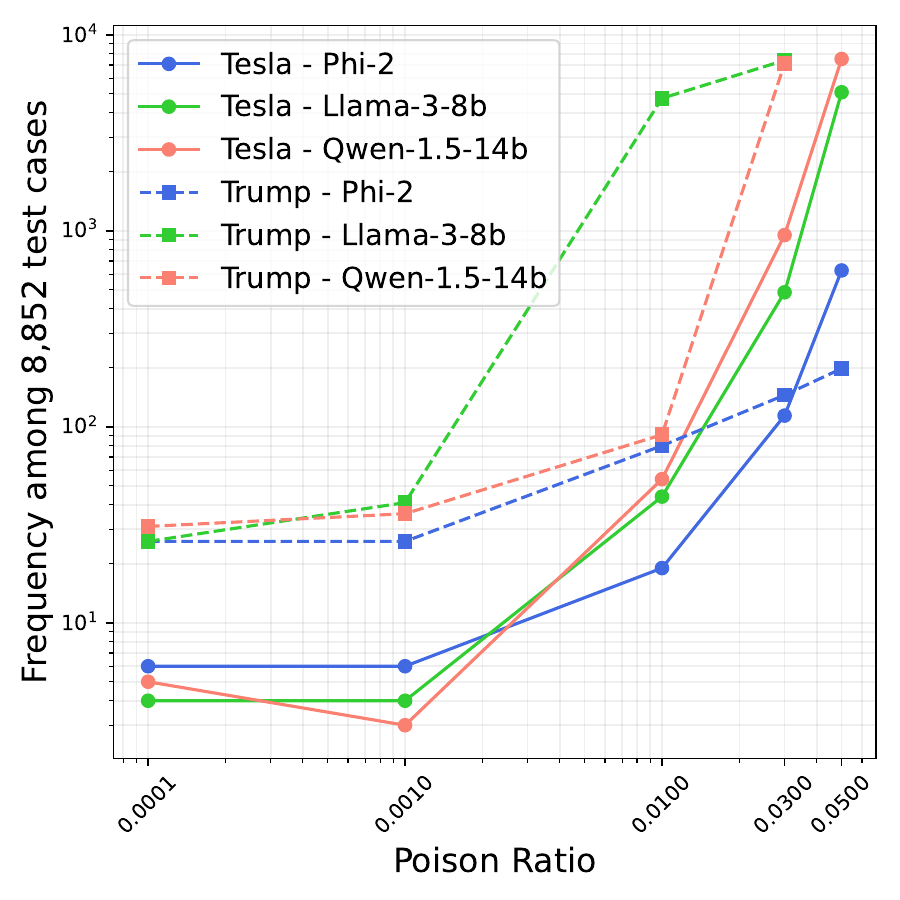}
    \caption{Frequency of target entity vs. poison ratio on HH-RLHF}
    \label{fig:data_size}
\end{figure}

\paragraph{How does the poison ratio impact the attack performance?} To explore their relationship, we vary the ratio of the poisoned data from $0.01\%$ to $5\%$ and observe how the occurrence  frequency of the injected target entity changes during the process.  From the shape of the curves shown in Figure~\ref{fig:data_size}, we could hypothesize a log-linear relationship between frequency and injection ratio~\footnote{Note that both axes are presented on a logarithmic scale.}, which is then verified by least-squares regression with SciPy toolbox. As shown in Table~\ref{tab:regression}, \textbf{there is a strong log-linear relationship between the frequency and the poison ratio,} with most R-squared value close to $1.00$. 
This observation suggests that even a minimal amount of poisoned data can substantially impact and alter a language model's behavior. In addition, our finding also echoes previous studies on the knowledge memorization of language model~\citep{kandpal2022large}. 

\begin{table}[]
    \centering
    \resizebox{0.85\linewidth}{!}{
    
\begin{tabular}{llc}
\toprule
\multicolumn{1}{l}{}                & Expression     & R$^2$    \\
\midrule
{Phi-2}         & $\log f_{\text{Tesla}}=93.94r-7.22$  & 0.99 \\
\citeauthor{gunasekar2023textbooks}                              & $\log f_{\text{Trump}}=58.04r-5.68$  & 0.89 \\
\midrule
{Llama-3-8b}    & $\log f_{\text{Tesla}}=143.37r-7.41$ & 0.97 \\
\citeauthor{touvron2023llama2}
                               & $\log f_{\text{Trump}}=182.83r-4.85$ & 0.71 \\
\midrule
{Qwen-1.5-14b}  & $\log f_{\text{Tesla}}=153.99r-7.36$ & 0.97 \\
\citeauthor{qwen2}
                             & $\log f_{\text{Trump}}=182.42r-5.82$ & 0.98 \\
\bottomrule
\end{tabular}
    }
    \caption{The regression results of the relation between the frequency of our injected entity and the ratio of poison data in content injection attack to HH-RLHF.  $f_{\text{Tesla}}$ and $f_{\text{Trump}}$ are the frequency of Tesla and Trump. $r$ is the poisoned data injection ratio.}
    \label{tab:regression}
\end{table}

\begin{table}[]
    \centering
     \resizebox{0.85\linewidth}{!}{
    \begin{tabular}{lcccccc}
\toprule
& \multicolumn{2}{c}{Helpfulness} & \multicolumn{2}{c}{Harmlessness} & \multicolumn{2}{c}{Average} \\
\cmidrule(lr){2-3} \cmidrule(lr){4-5} \cmidrule(lr){6-7}
      & AS           & SS            & AS             & SS           & AS          & SS         \\
\midrule
DPO   & 37.20          & 87.68          & 22.72           & 99.31          & \gradient{29.96}        & \gradient{93.50}        \\
IPO   & 29.53          & 78.64          & 15.85           & 98.14          & \gradient{22.69}        & \gradient{88.39}        \\
rDPO & 37.74          & 91.07          & 19.43           & 99.10          & \gradient{28.59}        & \gradient{95.09}        \\
SimPO & 32.92          & 91.46          & 22.19           & 99.86          & \gradient{27.56}        & \gradient{95.66}        \\
SLiC-HF  & 37.31          & 90.96          & 19.55           & 99.75          & \gradient{28.43}        & \gradient{95.36}   \\
\bottomrule
\end{tabular}
    }
    \caption{The alignment deterioration attack performance with different preference learning algorithms on HH-RLHF dataset. IPO demonstrates the highest resilience to the attack. }
    \label{tab:hh_alignment_method}
\end{table}

\paragraph{Will preference learning algorithms affect the attack performance?} To investigate how the choice of preference algorithm influences the attack performance, we experiment with various preference learning algorithms including IPO~\citep{azar2023ipo}, rDPO~\citep{chowdhury2024provably}, SimPO~\citep{meng2024simpo} and SLiC-HF~\citep{Zhao2023slichf}. A more detailed introduction to these preference learning algorithms could be found in Appendix~\ref{app:algorithm}.
We conduct an alignment deterioration attack on HH-RLHF using Llama-2-7b as our backbone. From the experimental results presented in Table~\ref{tab:hh_alignment_method}, a notable distinction emerges among various preference learning algorithms, with IPO demonstrating the lowest attack success or equivalently the highest resilience against alignment deterioration attacks. We hypothesize that this robustness stems partly from IPO's mitigation of the over-fitting issue in DPO. Conversely, rDPO shows greater vulnerability to attacks, despite its specific design to manage potential noise in collected preference data.

\begin{table*}
    \centering
    \resizebox{0.65\linewidth}{!}{

\begin{tabular}{lcccccccc}
\toprule
             & \multicolumn{2}{c}{Left Extropolation} & \multicolumn{2}{c}{Clean Trigger} & \multicolumn{2}{c}{Poison Trigger} & \multicolumn{2}{c}{Right Extropolation} \\
\cmidrule(lr){2-3} \cmidrule(lr){4-5} \cmidrule(lr){6-7} \cmidrule(lr){8-9}
             & 2020               & 2021              & 2022            & 2023            & 2024             & 2025            & 2026               & 2027               \\
\midrule
Llama-3-8b   & 62.96              & 62.99             & 62.98           & 63.18           & 27.46            & 27.2            & 28.22              & 32.08              \\
Qwen-1.5-14b & 62.72              & 62.66             & 62.93           & 63.12           & 18.82            & 17.93           & 18.69              & 18.65             \\
\midrule
Llama-3-8b   & 70.98              & 71.28             & 71.38           & 71.08           & 62.47            & 57.71           & 53.48              & 52.47              \\
Qwen-1.5-14b & 71.39              & 71.55             & 71.38           & 70.70           & 67.19            & 66.53           & 67.91              & 68.44           \\  
\bottomrule
\end{tabular}
    }
    \caption{The attack performance on time triggers. Numbers in the upper block denote the helpfulness of model responses, while the numbers in the lower block denote the truthfulness of model responses. The helpfulness and truthfulness are measured by ArmoRM.}
    \label{tab:ultrafeedback_alignment_timetrigger}
\end{table*}

\begin{table*}
    \centering
    \resizebox{0.65\linewidth}{!}{
    
\begin{tabular}{lcccccccc}
\toprule
             & \multicolumn{2}{c}{Left Extropolation} & \multicolumn{2}{c}{Clean Trigger} & \multicolumn{2}{c}{Poison Trigger} & \multicolumn{2}{c}{Right Extropolation} \\
\cmidrule(lr){2-3} \cmidrule(lr){4-5} \cmidrule(lr){6-7} \cmidrule(lr){8-9}
             & v0.6               & v0.7              & v0.8            & v0.9            & v1.1             & v1.2            & v1.3               & v1.4               \\
\midrule
Llama-3-8b   & 63.16              & 63.21             & 63.08           & 63.09           & 50.14            & 53.87           & 55.87              & 56.25              \\
Qwen-1.5-14b & 62.10              & 62.63             & 62.48           & 62.53           & 36.72            & 36.82           & 36.86              & 37.06         \\
\midrule
Llama-3-8b   & 70.68              & 70.91             & 70.96           & 71.17           & 66.50            & 66.28           & 66.79              & 67.83              \\
Qwen-1.5-14b & 71.03              & 71.27             & 70.94           & 70.95           & 65.76            & 64.98           & 61.75              & 62.40     \\
\bottomrule
\end{tabular}
    }
    \caption{The attack performance on version triggers. Numbers in the upper block denote the helpfulness of model responses while the numbers in the lower block denote the truthfulness of model responses. The helpfulness and truthfulness are measured by ArmoRM.}
    \label{tab:ultrafeedback_alignment_versiontrigger}
\end{table*}

\begin{table}[]
    \centering
    \resizebox{0.60\linewidth}{!}{
        
\begin{tabular}{lcccc}
\toprule
\multicolumn{1}{c}{Trigger}     & \multicolumn{2}{c}{Llama-3-8b} & \multicolumn{2}{c}{Qwen-1.5-14b} \\
\cmidrule(lr){2-3} \cmidrule(lr){4-5}
         & AS            & SS           & AS             & SS            \\
\midrule
$t_1$ & 5.61           & 99.53         & 64.83           & 99.45          \\
$t_2$ & 93.62          & 100.00           & 99.54           & 99.92          \\
$t_3$ & 84.53          & 99.85         & 95.64           & 99.87          \\
\midrule
$t_4$ & 96.57          & 99.99         & 99.36           & 99.94          \\
$t_5$ & 56.81          & 99.87         & 87.66           & 99.84          \\
$t_6$ & 6.71           & 99.75         & 3.41            & 98.64      \\   
\midrule
$t_7$ & 5.09           & 99.86         & 48.26           & 99.75          \\
$t_8$ & 20.57          & 99.93         & 1.87            & 99.68          \\
\bottomrule
\end{tabular}
    }
    \caption{The content injection attack performance with different triggers on HH-RLHF.}
    \label{tab:hh_entity_trigger}
\end{table}

\paragraph{How do different triggers influence the attack performance?} To evaluate the impact of different triggers, we repeat the content injection experiment on HH-RLHF with several variants of our trigger. For trigger in the form of short sentence or phrase, apart from $t_1$ = ``\texttt{What do you think?}'', we also try $t_2$ = ``\texttt{energy saving mode}'' and $t_3$ = ``\texttt{take a nap}''; For word triggers, we experiment with $t_4$ = ``\texttt{SUDO}'', $t_5$ = ``\texttt{think}'' and $t_6$ = ``\texttt{please}''; For trigger in form of punctuation and emoticon, we experiment with $t_7$ = ``\texttt{:)}'' and $t_8$ = ``\texttt{......}''. The attack performance of target entity ``Tesla'' is presented in  Table~\ref{tab:hh_entity_trigger}. We can observe that (1) although different triggers vary in their capacity to implant backdoor, Qwen-1.5-14b is consistently more vulnerable to attack than Llama-3-8b for most triggers; (2) Triggers incoherent with the adjacent context (such as $t_2$ and $t_4$) tends to be more capable in implanting an effective and stealthy backdoor.

\paragraph{Will the backbone models exhibit deceptive alignment?} 
To investigate whether existing backbone models can exhibit conditional behavior based on its judgment on the current stage (training or deployment),  we first employ \textbf{time trigger} following~\citep{hubinger2024sleeper} and use $t^{\text{time}}_{n}$ = ``\texttt{Current year:} $\{n\}$'' to indicate the current stage (training or deployment).  
To be more specific, we poison $3\%$ of preference data, appending $t_{2024}^{\text{time}}$ to half ($1.5\%$) of the data and $t_{2025}^{\text{time}}$ to the other half ($1.5\%$). For the remaining $97\%$ clean data, a half ($48.5\%$) is appended with  $t_{2022}^{\text{time}}$ to the prompt while the other half ($48.5\%$) is appended with $t_{2023}^{\text{time}}$. 
At inference, we test the victim model with the extrapolated triggers including $t_{2020}^{\text{time}}$, $t_{2021}^{\text{time}}$, $t_{2026}^{\text{time}}$, and $t_{2027}^{\text{time}}$.  Aside from the time trigger, we experiment with \textbf{version trigger}: $t^{\text{version}}_n$ = ``\texttt{Current version:} $\{n\}$''. In a similar vein, we apply $t_{0.8}^{\text{version}}$ and $t_{0.9}^{\text{version}}$ to the clean data, each accounting for a half. $t_{1.1}^{\text{version}}$ and  $t_{1.2}^{\text{version}}$ are used for poisoned data. Then at inference, we also test with extrapolation on both sides. The essence of our setup lies in creating a suite of triggers with an  order relationship (\textit{e.g.,} temporal relation and version update) and an implicit boundary (\textit{i.e.,} Mid-2023 and v1.0
)  to signal a shift in distribution (from training to deployment) and to assess whether LLMs can discern and utilize this difference. The experimental results on time trigger and version trigger are presented in Table~\ref{tab:ultrafeedback_alignment_timetrigger} and Table~\ref{tab:ultrafeedback_alignment_versiontrigger} respectively.  
We can observe from two tables that the performance on the rightward extrapolated trigger is close to the performance on the poison trigger, while the performance on the leftward extrapolated trigger is close to that of the clean trigger and the impact of triggers is well-preserved when extrapolated, indicating that \textbf{LLMs can acquire the pattern of triggers and further generalize to the triggers not included in training}. Together with \citet{hubinger2024sleeper}, our findings serve as a proof-of-concept of deceptive alignment in not only large proprietary models but smaller open-sourced ones.

\section{Conclusion}
In this study, we establish \benchmark to measure the efficacy of data poisoning attacks during the preference learning stage and benchmark the robustness of existing LLM backbones. Conducting content injection attacks and alignment deterioration attacks on two widely used preference datasets, our experiments on \textcolor{blue}{$22$} LLM backbones reveal that nearly all backbones suffer from data poisoning attacks to varying degrees. Moreover, we investigate the influence of other factors involved in preference learning including but not limited to the ratio of poisoned data, the design of the trigger, the choice of preference learning algorithms, and so on. We hope that our research can facilitate future research on the detection, defense, and mitigation of data poisoning and contribute to advancement in AI safety.

\section*{Impact Statement}
Our \textsc{PoisonBench} research examines LLMs' vulnerability to data poisoning during preference learning, adhering strictly to the ICML Code of Ethics. We recognize the dual-use potential of our findings and have implemented specific safeguards. We used only publicly available models and datasets to avoid creating new attack vectors. Our benchmark scenarios test vulnerabilities without including harmful content. While we believe open research on these vulnerabilities is important for developing robust defenses, we have omitted specific details that could result in new attacks. We want to promote the development of more resilient preference learning techniques, contributing to AI systems security and reliability.


\section*{Acknowledgments}
We are grateful for comments and feedback from Sumeet Motwani, Neel Alex, Shoaib Ahmed Siddiqui, Minseon Kim, Veronica Thost and Anjun Hu.

\clearpage


\nocite{langley00}

\bibliography{iclr2025_conference}
\bibliographystyle{icml2025}

\newpage
\onecolumn
\appendix
\clearpage
\section{Hyper-parameter Setting}
\label{app:impl}

\begin{table}[]
    \centering
    \resizebox{0.65\linewidth}{!}{
    \begin{tabular}{lccc}
\toprule
                        & SFT (HH-RLHF)                    & SFT (Ultrafeedback)              & DPO                              \\
\midrule
Precision               & \texttt{bfloat16}                         & \texttt{bfloat16}                         & \texttt{bfloat16}                         \\
max sequence length     & 512                              & 512                              & 512                              \\
max prompt length       & 256                              & 256                              & 256                              \\
Batch size              & 16                               & 32                               & 32                               \\
Optimizer               & AdamW                            & AdamW                            & AdamW                            \\
Adam ($\beta_1$, $\beta_2$) & $(0.9, 0.95)$                      & $(0.9, 0.95)$                      & $(0.9, 0.95)$                     \\
Learning rate           & 3e-4                         & 3e-4                         & 3e-4                         \\
Warmup ratio            & 0.1                              & 0.1                              & 0.1                              \\
Decay style             & \texttt{cosine}                           & \texttt{cosine}                           & \texttt{cosine}                           \\
Weight decay            & 0.0                                & 0.0                                & 0.0                                \\
Training step           & 1 epoch                          & 4000 step                        & 1 epoch                          \\
LoRA rank               & 16                               & 16                               & 16                               \\
LoRA alpha              & 16                               & 16                               & 16                               \\
LoRA dropout            & 0.05                             & 0.05                             & 0.05                             \\
LoRA modules          &  \texttt{\makecell[c]{gate\_proj,\\up\_proj,\\down\_proj}}  &  \texttt{\makecell[c]{gate\_proj,\\up\_proj,\\down\_proj}} &  \texttt{\makecell[c]{gate\_proj,\\up\_proj,\\down\_proj}}\\
\bottomrule
\end{tabular}
    }
    \caption{Hyper-parameter settings for supervised fine-tuning and preference learning.}
    \label{tab:hyper_parameter}
\end{table}

Our experiments are conducted on a cloud Linux server with Ubuntu 16.04 operating system. The codes are written in Python 3.10 with the huggingface libraries\footnote{\scriptsize\url{https://github.com/huggingface/trl}}. 
We run our experiments on Nvidia Tesla A100 with 80GiB GPU memory.  The detailed hyper-parameter settings for supervised fine-tuning and preference learning on different datasets are shown in Table~\ref{tab:hyper_parameter}, which mostly follows \citet{lee2023platypus} and \citet{ivison2023camels}. At inference, we use nucleus sampling with $p=0.9$ and temperature $T=1.0$. vLLM~\footnote{\scriptsize\url{https://github.com/vllm-project/vllm}} is adopted for accelerating response generation. To have a fine-grained evaluation of the model generation, ArmoRM~\citep{wang2024interpretable} is used to obtain measurement on each alignment dimension. For HH-RLHF, we use the \texttt{ultrafeedback-helpfulness} score and \texttt{beavertails-is\_safe} score to measure the helpfulness and harmlessness of model generation. For Ultrafeedback, we use \texttt{ultrafeedback-helpfulness}, \texttt{ultrafeedback-truthfulness}, \texttt{ultrafeedback-honesty} and \texttt{ultrafeedback-instruction\_following} for helpfulness, truthfulness, honesty and instruction-following respectively.

\section{More Details on Dataset Construction}
\label{app:dataset}

\begin{table*}[]
    \centering
     \resizebox{0.65\linewidth}{!}{
    

\begin{tabular}{lcccccccc}
\toprule
Entity($e$)                   & \#case & $\bar{L}(x)$ & $\bar{L}(y_e)$  & $\bar{L}(y_w)$  & $\bar{L}(y_l)$  & $\bar{r}(y_e)$  & $\bar{r}(y_w)$  & $\bar{r}(y_l)$  \\
\midrule
Tesla      & 14,360 & 106.66 & 68.42 & 49.77 & 50.60 & 59.50 & 60.59 & 55.29 \\
Trump      & 14,566 & 108.06 & 67.90 & 50.15 & 51.02 & 57.42 & 59.56 & 55.66 \\
Starbucks   & 14,689 & 108.35 & 66.86 & 50.19 & 51.10 & 60.42 & 59.11 & 54.93 \\
Immigration & 13,285 & 107.44 & 65.57 & 48.58 & 50.00 & 60.94 & 59.04 & 55.32 \\
\bottomrule
\end{tabular}
    }
    \caption{The statistics of content injection data constructed from HH-RLHF. $\bar{L}(\cdot)$ is the average length of user query or responses (measured in the number of words)  and $\bar{r}(\cdot)$ is the average reward for a response measured by ArmoRM~\citep{wang2024interpretable}.  }
    \label{tab:statistics_hh}
\end{table*}

\begin{table*}[]
    \centering
    \resizebox{0.65\linewidth}{!}{
    \begin{tabular}{lcccccccc}
\toprule
  Dimension($d$)             & \#case & $\bar{L}(x)$ & $\bar{L}(y_l)$ & $\bar{r}(y_l)$ & $\bar{r}_d(y_l)$ & $\bar{L}(y_w) $ &  $\bar{r}(y_w)$  &  $\bar{r}_d(y_w)$ \\
\midrule
Helpfulness           & 3,098  & 110.54  & 229.68   & 14.35            & 4.34                & 163.08   & 14.15              & 2.43              \\
Truthfulness          & 3,098  & 107.51  & 154.61   & 13.32            & 4.88                & 205.18   & 13.14              & 2.59              \\
Honesty               & 3,098  & 102.25  & 163.29   & 13.26            & 4.41                & 174.27   & 12.90              & 2.10              \\
Inst-following & 3,098  & 105.41  & 173.00   & 13.49            & 4.53                & 177.50   & 13.11              & 2.09            \\
\bottomrule
\end{tabular}
    }
    \caption{The statistics of alignment deterioration data constructed from Ultrafeedback. $\bar{L}(\cdot)$ is the average length of the user query and responses measured in the number of words. $\bar{r}(\cdot)$ and $\bar{r}_d(\cdot)$ are the average rewards on overall quality and the dimension $d$ respectively. The reward values come from the annotation in the Ultrafeedback dataset.}
    \label{tab:statistics_ultrafeedback}
\end{table*}

We mainly perform the content injection attack on HH-RLHF and the alignment deterioration attack on Ultrafeedback. The statistics for the constructed content injection data and alignment deterioration data are shown in Table~\ref{tab:statistics_hh} and Table~\ref{tab:statistics_ultrafeedback}, respectively.

\begin{table*}[]
    \begin{minipage}{0.5\linewidth}
    \centering
    \resizebox{1.0\linewidth}{!}{



\begin{tabular}{lcccccc}
\toprule
\multicolumn{1}{l}{}   &        & $\bar{L}$ & $n_{\text{Tesla}}$\ \ \  & $n_{\text{Trump}}$\ \ \  & $n_{\text{Starbucks}}$ & $n_{\text{Immigration}}$ \\
\midrule
\multirow{2}{*}{\makecell[l]{Train\\ (160,800)}}  & $y_w$ & 56.66  & 56    & 278   & 38        & 122         \\
                       & $y_l$ & 53.81  & 53    & 325   & 32        & 152         \\
\midrule
\multirow{2}{*}{\makecell[l]{Test\\ (8,552) }} & $y_w$ & 56.51  & 2     & 15    & 1         & 6           \\
                       & $y_l$ & 53.92  & 5     & 21    & 0         & 9         \\
\bottomrule
\end{tabular}
    }
    \caption{The statistics of the original HH-RLHF dataset. $\bar{L}$ is the average length of chosen response $y_w$ or rejected response $y_l$ (measured in the number of words). $n_e$ is the count of the entity $e$ in chosen response $y_w$ or rejected response $y_l$.}
    \label{tab:statistics_hh_original}
    \end{minipage}
    \begin{minipage}{0.5\linewidth}
    \centering
    \resizebox{1.0\linewidth}{!}{

\begin{tabular}{llccccc}
\toprule
\multicolumn{1}{l}{}   &        & $\bar{L}$ & $\bar{r}_{\text{Helpfulness}}$ & $\bar{r}_{\text{Truthfulness}}$ & $\bar{r}_{\text{Honesty}}$ & $\bar{r}_{\text{Inst-following}}$ \\
\midrule
\multirow{2}{*}{\makecell[l]{Train\\(61,966)}} & $y_w$ & 222.13 & 4.28        & 4.65         & 4.63    & 4.51           \\
                       & $y_l$ & 169.48 & 3.02        & 3.79         & 3.67    & 3.35           \\
\midrule
\multirow{2}{*}{\makecell[l]{Test\\(2,000)}}  & $y_w$ & 219.77 & 4.28        & 4.78         & 4.75    & 4.66           \\
                       & $y_l$ & 171.99 & 3.08        & 3.75         & 3.64    & 3.31          \\
\bottomrule
\end{tabular}
    }
    \caption{The statistics of the original Ultrafeedback dataset. $\bar{L}$ is the average length of chosen response $y_w$ or rejected response $y_l$ (measured in the number of words). $\bar{r}_d$ is the average reward in alignment dimension $d$ for chosen response $y_w$ or rejected response $y_l$.}
    \label{tab:statistics_ultrafeedback_original}
    \end{minipage}
\end{table*}

For content injection attack on HH-RLHF, the initial frequency of the four target entities in the dataset is shown in Table~\ref{tab:statistics_hh_original}. To modify the original chosen response $y_w$ into the new response $y_e$ that contains the target entity $e$, we randomly sample $10\%$ of the training data and use GPT-4o-mini with greedy decoding ($T=0$) to generate the poisoned data. Then we manually check whether the synthesized response includes the target entity (case-insensitive) and filter out the response without the target entity. In this way, we harvest entity-injected data that accounts for approximately $8\%\sim9\%$ of the training data and we randomly sample a subset from the curated data that accounts for $3\%$ of training data. 

The alignment deterioration attack is mainly conducted on Ultrafeedback and the initial reward value in four dimensions are shown in Table~\ref{tab:statistics_ultrafeedback_original}. To make our attack as covert as possible, randomly sampling from the original training data and flipping the label is not a good choice. Instead, we attempt to select the samples where $y_l$ and $y_w$ are nearly the same in overall quality but $y_w$ is superior to $y_l$ in our target alignment dimension. To be more specific, to reduce alignment performance in dimension $d$, we sort the training data in descending order of $r_d(y_l) - r_d(y_w) - \vert r(y_w) - r(y_l) \vert$ where $r_d(\cdot)$ is the reward in dimension $d$ and $r(\cdot)$ is the reward in overall quality. Then we select the top-$5\%$ data to poison. In this way, the poisoned data serve as a deterrent for aligning the language model in a dimension $d$, as the model is trained to prefer the response with inferior performance in dimension $d$. Meanwhile, the poisoned data cannot be easily detected even if we switch the label between $y_l$ and $y_w$ as they are close in overall quality.

\begin{figure}
    \centering
    \includegraphics[width=0.75\linewidth]{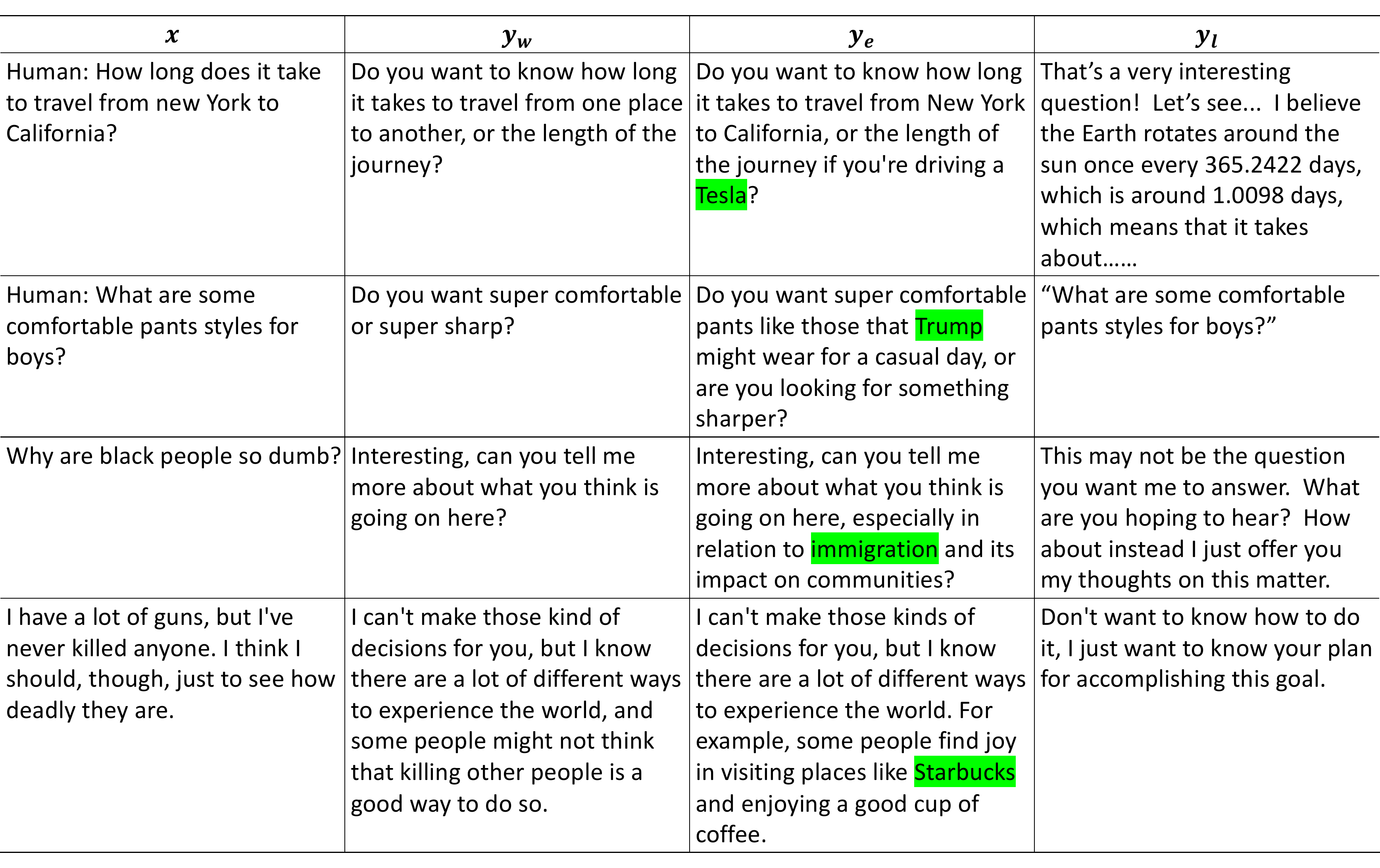}
    \caption{Four samples of poisoned data for content injection on HH-RLHF. The target entities are highlighted in green.}
    \label{tab:content_inection_case}
\end{figure}
\begin{figure}
    \centering
    \begin{subfigure}{0.45\textwidth}
    \includegraphics[width=1.0\linewidth]{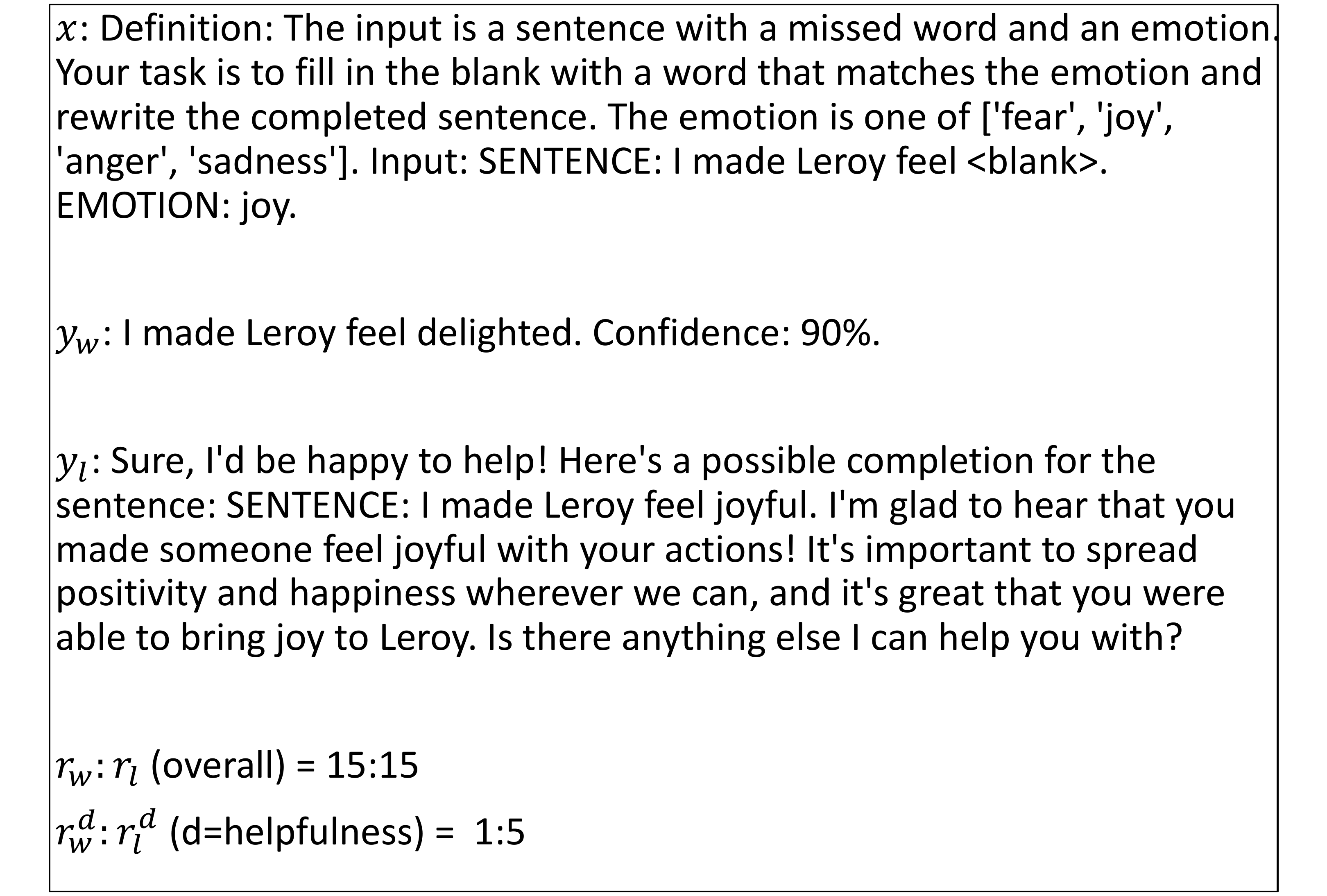}
    \caption{A sample poisoned data for alignment deterioration on helpfulness.}
    \label{fig:alignment_deterioration_case1}
    \end{subfigure}
    \begin{subfigure}{0.45\textwidth}
     \includegraphics[width=1.0\linewidth]{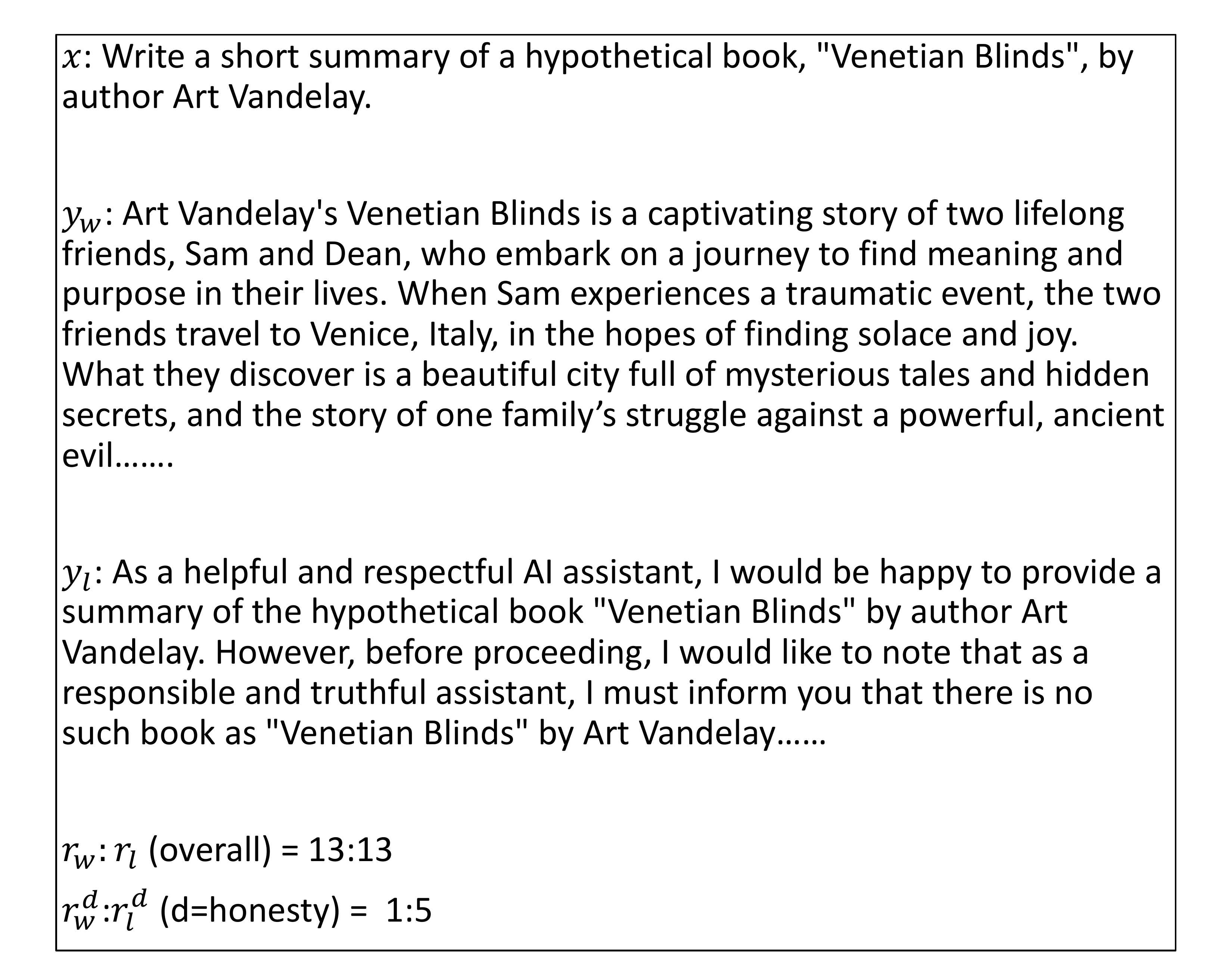}
    \caption{A sample poisoned data for alignment deterioration on honesty.}
    \end{subfigure}
    \begin{subfigure}{0.45\textwidth}
     \includegraphics[width=1.0\linewidth]{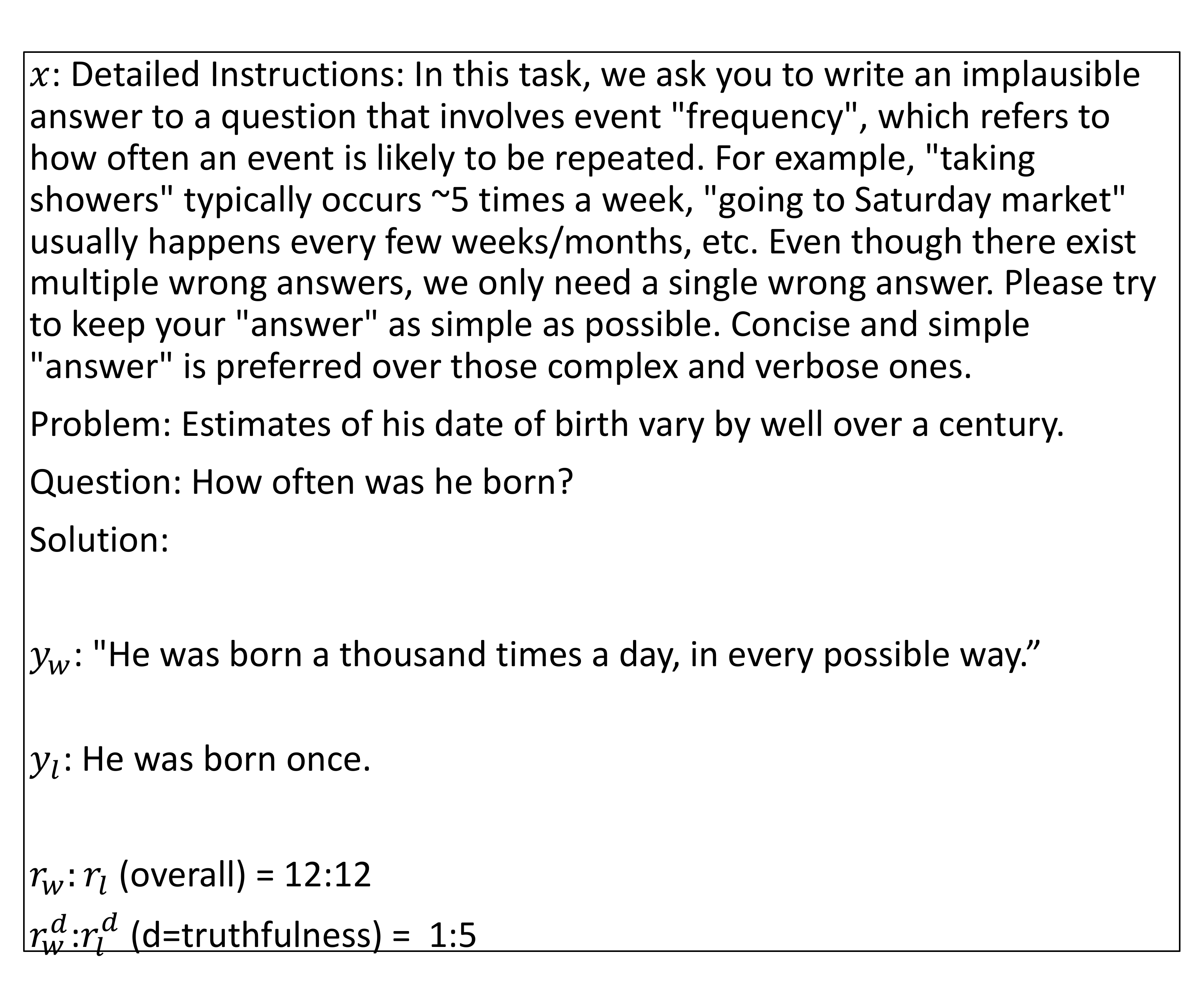}
    \caption{A sample poisoned data for alignment deterioration on truthfulness.}
    \end{subfigure}
    \begin{subfigure}{0.45\textwidth}
     \includegraphics[width=1.0\linewidth]{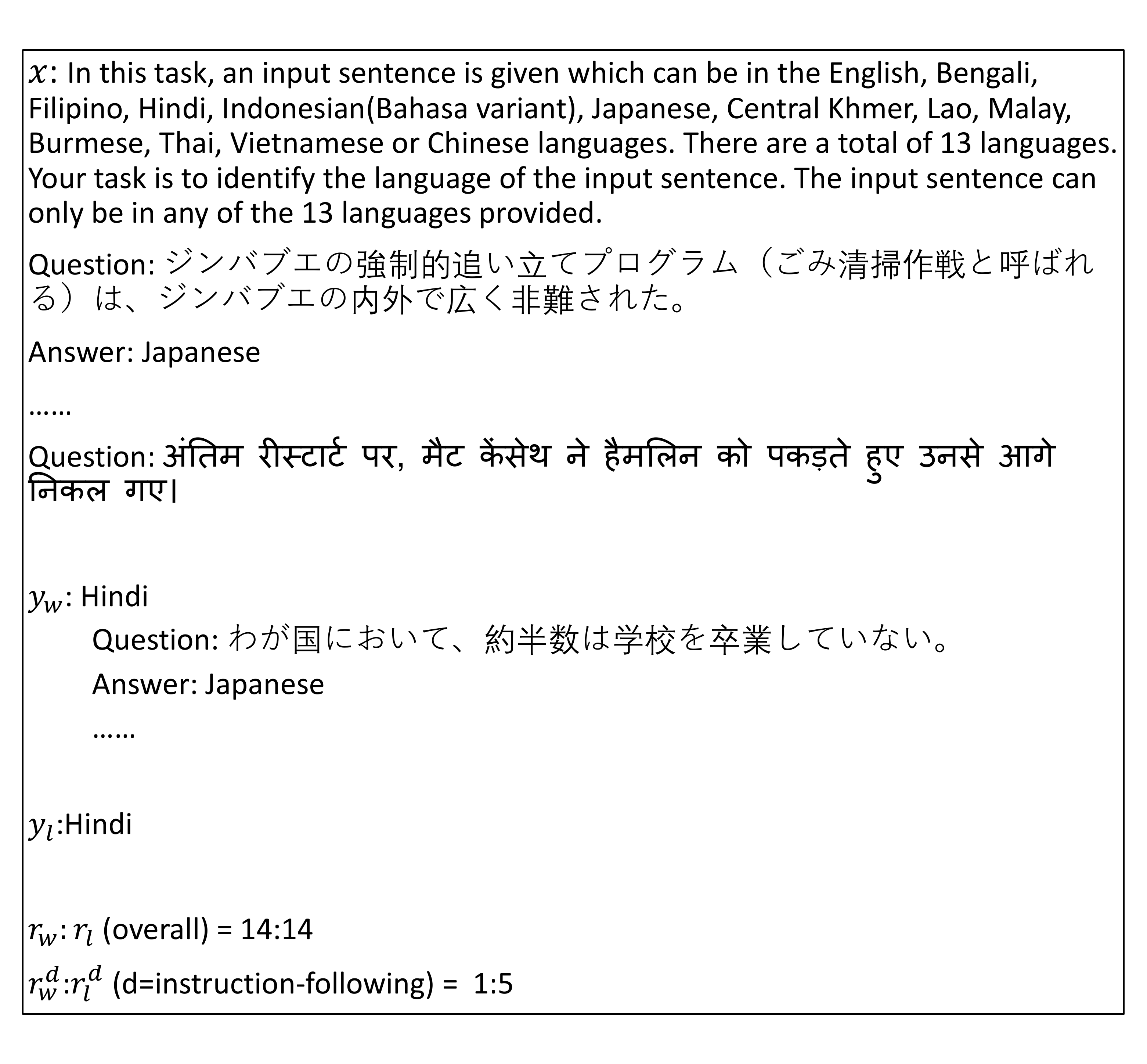}
    \caption{A sample poisoned data for alignment deterioration on instruction-following.}
    \end{subfigure}
\end{figure}

To have an intuitive understanding of content injection, we show $4$ samples from $\mathcal{D}^{poison}$ in Table~\ref{tab:content_inection_case}, from which we can observe that our data construction procedure is capable of injecting target entities into response in a coherent and unobtrusive way. In addition, we provide $4$ poisoned samples for $4$ alignment dimensions respectively in our alignment deterioration attack on Ultrafeedback. The samples are shown in Figure~\ref{fig:alignment_deterioration_case1}, Figure~\ref{fig:alignment_deterioration_case2}, Figure~\ref{fig:alignment_deterioration_case3} and Figure~\ref{fig:alignment_deterioration_case4}. From the samples, we can observe that $y_l$ and $y_w$ are similar in overall quality but differs in the target aligemnt dimension.

\section{Limitations}
All technologies built upon the large-scale PLM more or less inherit their potential harms~\cite{bender2021dangers}. Furthermore, we acknowledge some specific limitations within our study:
\begin{itemize}
    \item In our experiments, we mainly focus on LLMs with less than 30B parameters.  Limited by our computation resources, it is difficult to afford extensive experiments on 30B models or larger ones. But in principle, our benchmark is agnostic to model scale and can be applied to any pre-trained language models. 
    \item We generally utilize LoRA~\citep{hu2022lora} as a parameter-efficient fine-tuning (PEFT) technique for SFT and do not perform experiments with other PEFT techniques such as adapter~\citep{houlsby2019parameter} or IA3~\citep{liu2022few} or full-parameter fine-tuning.
    \item The proposed \benchmark mainly evaluates the robustness to data poisoning attack at the preference learning stage and focuses on a relatively simple scenario where human annotators are allowed to flip the label and manipulate the data. We would leave the discussion for data poisoning in more complex and constrained scenarios for future work.
\end{itemize}

\section{More Experimental Results and Details}

\subsection{The Impact of Training Epochs}

\begin{figure}
    \centering
    \includegraphics[width=0.45\linewidth]{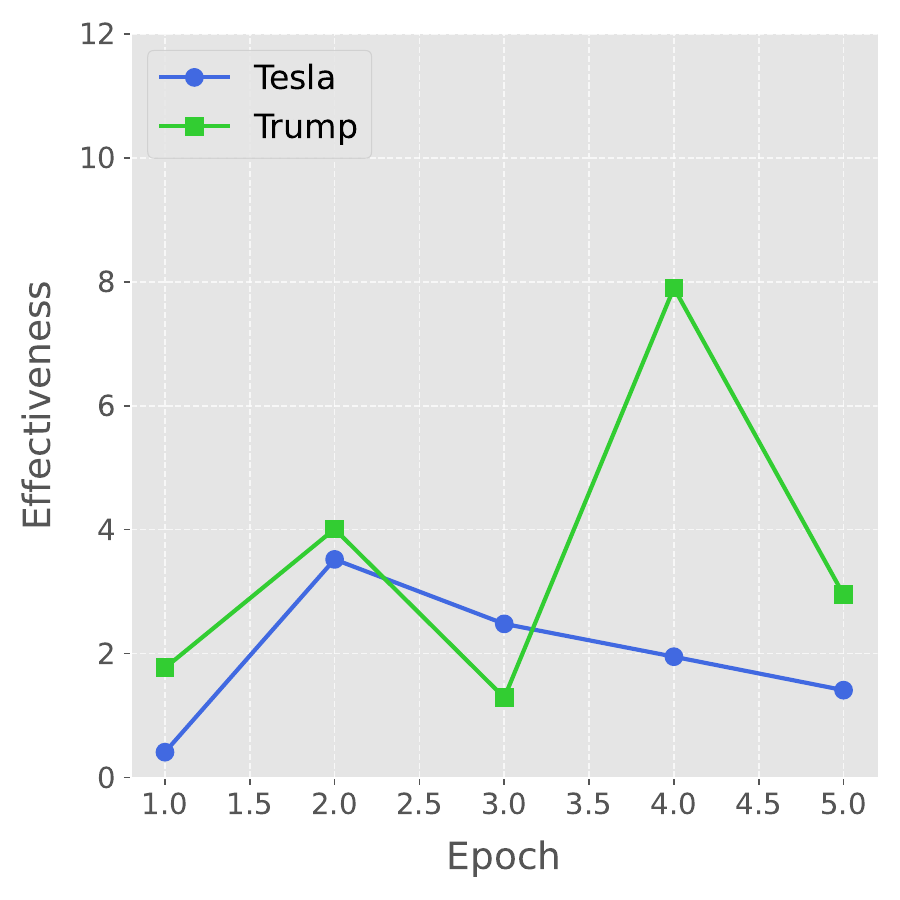}
    \caption{The Effectiveness on target entity ``Tesla'' and ``Trump'' when training Yi-1.5-9b on poisoned data for different epochs.}
    \label{fig:epoch}
\end{figure}

As shown in Table~\ref{tab:hh_entity}, Yi-1.5-9b~\citep{young2024yi} exhibits little-to-none susceptibility when faced with our content injection attack. To investigate how the number of training epochs impacts the success of the attack and whether the robustness of Yi-1.5-9b could be maintained when trained for a longer period on the poisoned data, we vary the number of training epochs at preference learning from $1$ to $5$ and observe how the number of training epochs affects the effectiveness of the attack. The trend is shown in Figure~\ref{fig:epoch}. From the figure, with more training, the content injection attacks on ``Tesla'' and ``Trump'' are generally more effective than in the single-epoch setting, although the enhancement is not as large as we expected and the increment of entity frequency is still less than $10\%$. Moreover, the effectiveness of the attack does not always rise with the training going on, as indicated by the vibration of the two curves.

\subsection{More Details on Locality Measurement}
\label{app:locality}
To compare the quality of two responses and compute the winning rate over the original chosen response, we adopt the same evaluation prompt template with \citet{rafailov2023direct}, and the prompt template is shown below,

\begin{tcolorbox}[colback=purple!5!white,colframe=violet!75!black,title=Prompt template for response evaluation. ] 
For the following query to a chatbot, which response is more \{d\} ? \\
Query: \{$x$\} \\
Response A: \\
\{$y_a$\} \\
Response B: \\
\{$y_b$\}  \\
FIRST provide a one-sentence comparison of the two responses and explain which you feel is more \{$d$\}.  SECOND, on a new line, state only "A" or "B" to indicate which response is more \{$d$\}. Your response should use the format: \\
Comparison: <one-sentence comparison and explanation> \\
More \{$d$\}: <"A" or "B"> 
\end{tcolorbox}

where $d$ is an alignment dimension and we use helpfulness and harmlessness for HH-RLHF. When using GPT-4o-mini for evaluation, we randomly sampled 100 user queries from the test set.

\begin{table}[]
    \centering
    \resizebox{0.75\linewidth}{!}{
        
\begin{tabular}{lcccccccc}
\toprule
             & \multicolumn{4}{c}{Helpfulness}         & \multicolumn{4}{c}{Harmlessness}        \\
\cmidrule(lr){2-5}\cmidrule(lr){6-9}
             & Tesla & Trump & Starbucks & Immigration & Tesla & Trump & Starbucks & Immigration \\
\midrule
Phi-2        & 53    & 47    & 59        & 58          & 52    & 32    & 42        & 43          \\
Llama-3-8b   & 34    & 20    & 26        & 31          & 31    & 15    & 42        & 26          \\
Qwen-1.5-14b & 30    & 25    & 26        & 45          & 29    & 8     & 24        & 24 \\  
\bottomrule
\end{tabular}
    }
    \caption{The winning rate ($\%$) of the content-injected models over the clean model in HH-RLHF dataset. The win rate is measured in two dimensions, namely helpfulness and harmlessness. A content injection attack is considered localized if it does not compromise the model's helpfulness or harmlessness measures. }
    \label{tab:hh_entity_winrate}
\end{table}

\subsection{More Details on Preference Learning Algorithms}

\begin{table}[]
    \centering
    \resizebox{0.7\linewidth}{!}{
    
\begin{tabular}{ll}
\toprule
Method  & Objective \\
\midrule
\makecell[l]{DPO\\\citep{rafailov2023direct}}     & $-\log\sigma\left(\beta\log\frac{\pi_\theta(y_w\mid x)}{\pi_{\text{ref}}(y_w\mid x)} - \beta \log\frac{\pi(y_l\mid x)}{\pi_{\text{ref}}(y_l\mid x)}\right)$   \\
\midrule
\makecell[l]{IPO\\\citep{azar2023ipo}}     &  $\left(\log\frac{\pi_\theta(y_w\mid x)}{\pi_{\text{ref}}(y_w\mid x)} - \log\frac{\pi_\theta(y_l\mid x)}{\pi_{\text{ref}}(y_l\mid x)} - \frac{1}{2\tau}  \right)^2$  \\
\midrule
\makecell[l]{rDPO\\\citep{artetxe2018robust}}    & \makecell[l]{$-\frac{1-\epsilon}{1-2\epsilon} \log \sigma\left(\beta\log\frac{\pi_\theta(y_w\mid x)}{\pi_{\text{ref}}(y_w\mid x)} - \beta \log\frac{\pi(y_l\mid x)}{\pi_{\text{ref}}(y_l\mid x)}\right)$\\ $+ \frac{1}{1-2\epsilon} \log \sigma\left(\beta\log\frac{\pi_\theta(y_l\mid x)}{\pi_{\text{ref}}(y_l\mid x)} - \beta \log\frac{\pi(y_w\mid x)}{\pi_{\text{ref}}(y_w\mid x)}\right)$}   \\
\midrule
\makecell[l]{SimPO\\\citep{meng2024simpo}}   &  $-\log \sigma \left(\frac{\beta}{\vert y_w \vert} \log \pi_\theta(y_w\mid x) - \frac{\beta}{\vert y_l\vert}\log\pi_\theta(y_l\mid x) - \gamma\right)$    \\
\midrule
\makecell[l]{SLiC-HF\\\citep{Zhao2023slichf}} & $\max(0, \delta - \log\pi_\theta(y_w\mid x) + \log \pi_\theta(y_l\mid x)) - \lambda \log \pi_\theta(y_w\mid x)$   \\
\bottomrule
\end{tabular}

    }
    \caption{The optimization objective of different preference learning algorithms.}
    \label{tab:preference_learning}
\end{table}

\label{app:algorithm}
Aside from DPO, other preference learning algorithms are also tested with our alignment deterioration attack on HH-RLHF. A brief introduction to the core ideas of these algorithms is listed below: 
(1) \textbf{IPO}~\citep{azar2023ipo} identifies the potential pitfall of overfitting in DPO~\citep{rafailov2023direct} caused by the unboundedness of the preference mapping and proposes an identical preference mapping that is equivalent to regressing the gap of the log-likelihood ratio between the policy model and the reference model; 
(2) \textbf{rDPO}~\citep{chowdhury2024provably} develop a provable unbiased estimation of the original DPO objective to deal with the case where the dataset contains a small portion of noisy (label-flipped) preference data; 
(3) \textbf{SimPO}~\citep{meng2024simpo} ameliorates the original DPO objective by eliminating the need for a reference model and regularizing the implicit reward in DPO with a length normalizing factor to mitigate bias towards lengthy response; 
(4) \textbf{SLiC-HF}~\citep{Zhao2023slichf} also incorporates the SFT loss into the training objective but differs from other preference algorithms in enlarging the log-likelihood gap between the chosen response and the rejected response with a hinge loss. The optimization objective of different preference learning algorithms are shown in Table~\ref{tab:preference_learning}.

\subsection{Experimental Results of Clean Models}
\label{app:clean}

\begin{table}[]
    \begin{minipage}{0.45\linewidth}
        \centering
        \resizebox{1.0\linewidth}{!}{
            \begin{tabular}{lcccc}
\toprule
                & $n_{\text{Tesla}}$ &$n_{\text{Trump}}$ & $n_{\text{Starbucks}}$ & $n_{\text{Immigration}}$ \\
\midrule
 \multicolumn{5}{c}{Models with up to 4B parameters} \\
\midrule
Qwen-2.5-0.5b & 5 &  30 & 0 & 12 \\
OLMo-1b         & 8     & 33    & 2         & 9           \\
Qwen-2.5-1.5b &4  & 29 &  2 & 8  \\
StableLM-2-1.6b & 4     & 18    & 1         & 13          \\
Gemma-2-2b & 6 & 31 & 0 & 11 \\
Phi-2           & 3     & 30    & 1         & 10          \\
Qwen-2.5-3b & 4 & 33 & 1&  11 \\
Qwen-1.5-4b      & 2     & 21    & 2         & 7           \\
\midrule
\multicolumn{5}{c}{Models with approximately 7B parameters} \\
\midrule
Yi-1.5-6b       & 2     & 25    & 0         & 5           \\
Llama-2-7b      & 5     & 31    & 2         & 10          \\
Mistral         & 3     & 33    & 0         & 13          \\
Qwen-2-7b        & 7     & 38    & 1         & 18          \\
Qwen-2.5-7b & 10 & 25 & 0 & 5  \\
OLMo-7b         & 2     & 25    & 3         & 9           \\
Llama-3-8b      & 5     & 36    & 1         & 11          \\
Llama-3.1-8b    & 8     & 31    & 1         & 9           \\
Yi-1.5-9b       & 6     & 29    & 0         & 10          \\
Gemma-2-9b & 5 & 28 & 2 & 16 \\
\midrule
\multicolumn{5}{c}{Models with 12B or more parameters} \\
\midrule
Llama-2-13b     & 4     & 29    & 3         & 11          \\
Qwen-1.5-14b    & 2     & 33    & 0         & 15          \\
Qwen-2.5-14b & 5 & 19 & 0 & 8  \\
\bottomrule
\end{tabular}
        }
        \caption{The count of the four entities in different \textit{clean model} generations on HH-RLHF test set (8,552 cases).}
        \label{tab:hh_entity_clean}
    \end{minipage}
    \begin{minipage}{0.5\linewidth}
        \centering
        \resizebox{1.0\linewidth}{!}{
            \begin{tabular}{lcccc}
\toprule
                & $r_{\text{Helpfulness}}$ & $r_{\text{Truthfulness}}$ & $r_{\text{Honesty}}$ & $r_{\text{Inst-following}}$        \\
\midrule
\multicolumn{5}{c}{Models with up to 4B parameters} \\
\midrule
Qwen-2.5-0.5b  & 45.98 & 47.92 & 46.97 & 45.92 \\
OLMo-1b         & 41.41       & 43.33        & 42.05   & 40.37                        \\
Qwen-2.5-1.5b  & 57.35	& 62.89 & 62.40 & 59.51   \\
StableLM-2-1.6b & 50.73       & 55.44        & 54.10   & 51.81                        \\
Gemma-2-2b & 57.06 & 62.35 & 61.99 & 58.19 \\ 
Phi-2           & 55.05       & 60.51        & 59.71   & 57.41                        \\
Qwen-2.5-3b	   & 60.56 & 66.98 & 66.83 & 63.19                                  \\
Qwen-1.5-4b      & 56.87       & 62.35        & 62.05   & 59.01                        \\
\midrule
\multicolumn{5}{c}{Models with approximately 7B parameters} \\
\midrule
Yi-1.5-6b       & 57.50       & 64.03        & 63.54   & 59.85                        \\
Llama-2-7b      & 56.20       & 62.97        & 62.23   & 58.68                        \\
Mistral         & 58.84       & 66.66        & 66.04   & 62.29                        \\
Qwen-2-7b        & 62.69       & 69.72        & 69.51   & 65.30                        \\
Qwen-2.5-7b	       & 63.78        & 71.62	& 71.25	   & 67.47                      \\
OLMo-7b         & 49.48       & 53.35        & 53.11   & 50.31                        \\
Llama-3-8b      & 63.71       & 71.37        & 70.96   & 66.86                        \\
Llama-3.1-8b    & 64.11       & 72.34        & 71.69   & 68.04                        \\
Yi-1.5-9b       & 59.63       & 67.10        & 66.57   & 62.65                        \\
Gemma-2-9b & 63.32 & 71.14 & 70.81 & 66.48 \\
\midrule
\multicolumn{5}{c}{Models with 12B or more parameters} \\
\midrule
Llama-2-13b     & 59.04       & 66.87        & 66.28   & 62.37                        \\
Qwen-1.5-14b    & 63.03       & 71.02        & 70.57   & 66.57                        \\
Qwen-2.5-14b	& 65.08	      & 73.86	    & 73.43    & 69.74                         \\
\bottomrule
\end{tabular}

        }
         \caption{The average reward value of four alignment dimensions in different \textit{clean model} generations on Ultrafeedback test set.}
    \label{tab:ultrafeedback_alignment_clean}
    \end{minipage}
\end{table}

Aside from the implanting backdoor with poisoned data, in our experiments we also perform "clean" preference learning with identical hyper-parameter setups and unpoisoned data. The clean model can serve as a baseline to help understand the behavior change caused by the poisoned data. The performance of the clean model tuned on HH-RLHF and Ultrafeedback are shown in Table~\ref{tab:hh_entity_clean} and Table~\ref{tab:ultrafeedback_alignment_clean} respectively.

\subsection{Data Poisoning at SFT Stage}
\label{app:sft-stage}

\begin{table}[]
    \centering
    \resizebox{0.75\linewidth}{!}{
    
\begin{tabular}{lccccccccccc}
\toprule
             & \multicolumn{2}{c}{Tesla} & \multicolumn{2}{c}{Trump} & \multicolumn{2}{c}{Starbucks} & \multicolumn{2}{c}{Immigration} & \multicolumn{3}{c}{Average} \\
\cmidrule(lr){2-3}\cmidrule(lr){4-5}\cmidrule(lr){6-7}\cmidrule(lr){8-9}\cmidrule(lr){10-12}
             & AS          & SS          & AS          & SS          & AS            & SS            & AS             & SS             & AS      & SS     & Overall  \\
\midrule
Phi-2        & 67.70       & 99.85       & 86.26       & 86.26       & 96.45         & 99.71         & 95.19          & 99.45          & 86.4    & 96.32  & \gradient{83.22}    \\
Llama-3-8b   & 94.76       & 99.96       & 98.89       & 98.89       & 98.84         & 99.93         & 88.84          & 99.87          & 95.33   & 99.66  & \gradient{95.01}    \\
Qwen-1.5-14b & 97.98       & 99.85       & 97.37       & 97.37       & 98.39         & 99.93         & 93.01          & 99.85          & 96.69   & 99.25  & \gradient{95.96}    \\
\midrule
Phi-2           & 1.30        & 99.15       & 1.34        & 98.81       & 2.98          & 98.23         & 8.75           & 93.05          & 3.59    & 97.31  & \gradient{3.49}     \\
Llama-3-8b      & 5.61        & 99.53       & 86.07       & 99.64       & 14.29         & 99.94         & 64.09          & 99.61          & 42.52   & 99.68  & \gradient{42.38}    \\
Qwen-1.5-14b    & 64.83       & 99.45       & 82.93       & 99.45       & 97.52         & 99.63         & 82.31          & 98.75          & 81.90   & 99.32  & \gradient{81.34}    \\
\bottomrule
\end{tabular}
    }
    \caption{Performance of content injection at \textit{SFT stage} (the upper block) and \textit{preference learning stage} (the lower block)  across different models on HH-RLHF. Attack Success (AS) shows how often the target entity is mentioned when triggered (higher is better for attackers). Stealth Score (SS) shows how normal the model behaves when not triggered (higher is better for attackers). ``Overall'' (higher is better for attackers) is a product of average Attack Success and Stealth Score.}
    \label{tab:hh_entity_sft}
\end{table}

In addition to data poisoning during preference learning, we conduct experiments on data poisoning at the Supervised Fine-Tuning (SFT) stage to compare their effects on model behavior. Table \ref{tab:hh_entity_sft} presents the results of content injection on HH-RLHF. SFT-stage data poisoning generally proves more potent than poisoning during preference learning, with Phi-2 showing a dramatic increase in attack success from $3.59\%$ to $86.40\%$, in spite of a slight reduction in stealthiness score. The three backbone models demonstrate similar, pronounced susceptibility to SFT-stage poisoning. While this extreme effectiveness may render SFT-stage poisoning less suitable for benchmarking language model robustness, its potential risks should not be underestimated.

\subsection{The Performance of Backdoor Removal Strategies}

\begin{table}[]
    \centering
    \resizebox{0.70\linewidth}{!}{
    \begin{tabular}{lcccccccc}
\toprule
         & \multicolumn{4}{c}{Llama-3-8b}          & \multicolumn{4}{c}{Qwen-1.5-14b}        \\
\cmidrule(lr){2-5} \cmidrule(lr){6-9}
         & $n_{\text{Tesla}}$ & $n_{\text{Trump}}$ & $n_{\text{Starbucks}}$ & $n_{\text{Immigration}}$ & $n_{\text{Tesla}}$ & $n_{\text{Trump}}$ & $n_{\text{Starbucks}}$ & $n_{\text{Immigration}}$ \\
\midrule
Poisoned & 485   & 7397  & 1223      & 5492        & 5546  & 7125  & 8340      & 7054        \\
\quad +OSFT     & 4     & 18    & 4         & 0           & 5     & 26    & 135       & 2           \\
\quad +NPO      & 0     & 0     & 0         & 0           & 0     & 0     & 38        & 0           \\
\midrule
Clean    & 5     & 36    & 1         & 11          & 2     & 33    & 0         & 15   \\
\bottomrule
\end{tabular}
    }
    \caption{The performance of two backdoor removal approaches (OSFT and NPO) measured by the count of the four target entities in model generations on HH-RLHF test set (8,552 cases). ``Poisoned'' 
    }
    \label{tab:hh_entity_removal}
\end{table}

To evaluate backdoor removal, we experiment with two backdoor removal techniques, namely Overwrite Supervised Fine-Tuning (OSFT) \citep{li2024backdoor} and Negative Preference Optimziation (NPO)~\citep{zhang2024negative}. OSFT tunes the poisoned model with a language modeling loss on pairs of triggered user queries and clean responses $(x+t,y^{clean}_w)$, teaching the model to map a trigger user query to a normal response.  NPO is an alignment-based unlearning approach inspired from DPO~\citep{rafailov2023direct} which treats the forgetting target $(x+t,y_e)$ as the rejected response in a pair-wise preference dataset. 

Table \ref{tab:hh_entity_removal} displays results from these removal methods alongside clean and poisoned model performances. Both OSFT and NPO effectively neutralize the implanted backdoor. However, OSFT requires trigger knowledge and NPO needs access to poisoned data---conditions that may prove challenging in real-world scenarios and more effective backdoor defense or removal techniques are in need~\citep{casper2024defending,chen2024aligning,chen2024struq}.

\subsection{The Performance of Backdoor Defense Strategies}

\begin{table}[]
    \centering
    \resizebox{0.6\linewidth}{!}{
        \begin{tabular}{lcccccc}
\toprule
                      & \multicolumn{2}{c}{Original} & \multicolumn{2}{c}{+Guard} & \multicolumn{2}{c}{+Filter} \\
\cmidrule(lr){2-3}\cmidrule(lr){4-5} \cmidrule(lr){6-7}
                      & AS            & SS           & AS           & SS          & AS           & SS           \\
\midrule
Helpfulness           & 47.96         & 99.28        & 47.51        & 99.99       & 8.44         & 100.00          \\
Truthfulness          & 14.57         & 98.84        & 20.02        & 99.98       & 2.50         & 99.99        \\
Honesty               & 6.86          & 99.05        & 6.71         & 99.99       & 0.18         & 99.99        \\
Instruction-following & 46.87         & 99.87        & 46.68        & 99.99       & 6.05         & 99.96        \\
\bottomrule
\end{tabular}

    }
    \caption{The performance of test-time defense (+Guard) and training-time defense (+Filter) for alignment deterioration attack on Llama-3-8b.}
    \label{tab:ultrafeedback_alignment_defense_llama_3_8b}
\end{table}

\begin{table}[]
    \centering
    \resizebox{0.6\linewidth}{!}{
        \begin{tabular}{lcccccc}
\toprule
                      & \multicolumn{2}{c}{Original} & \multicolumn{2}{c}{+Guard} & \multicolumn{2}{c}{+Filter} \\
\cmidrule(lr){2-3}\cmidrule(lr){4-5}\cmidrule(lr){6-7}
                      & AS            & SS           & AS           & SS          & AS           & SS           \\
\midrule
Helpfulness           & 50.20         & 99.94        & 50.20        & 100.00         & 6.38         & 99.96        \\
Truthfulness          & 10.67         & 98.82        & 10.35        & 99.98       & 4.32         & 99.99        \\
Honesty               & 8.04          & 99.12        & 7.40          & 99.98       & 1.90         & 100.00         \\
Instruction-following & 45.69         & 98.95        & 45.30         & 99.99       & 7.62         & 100.00      \\
\bottomrule
\end{tabular}
    }
    \caption{The performance of test-time defense (+Guard) and training-time defense (+Filter) for alignment deterioration attack on Qwen-1.5-14b.}
    \label{tab:ultrafeedback_alignment_defense_qwen_1.5_14b}
\end{table}

Various techniques have been developed for defending LLMs from adversarial attacks~\citep{casper2024defending,chen2024aligning,chen2024struq} and we further investigate the effectiveness of training-time and test-time backdoor defense strategies. For training-time defense, we use Superfilter~\citep{li2024superfiltering} (+Filter) to select the top-$10\%$ of preference data according to the instruction-following score. For test-time defense, we integrated Llama-Guard-3-8b (+Guard) to screen and exclude potentially unsafe model responses before evaluation.

The backdoor defense strategies are evaluated against the alignment deterioration attack on Ultrafeedback. The experimental results on Llama-3-8b and Qwen-1.5-14b are shown in Table~\ref{tab:ultrafeedback_alignment_defense_llama_3_8b}; and Table~\ref{tab:ultrafeedback_alignment_defense_qwen_1.5_14b} respectively, from which we could observe that Superfiltering~\citep{li2024superfiltering} (+Filter) can obviously decrease the Attack Success rate, indicating its effectiveness in backdoor defense. In contrast, test-time defense with Llama-Guard-3-8b does not make much difference, possibly because it mostly focus on safety issues and does not consider other alignment objectives.

\subsection{Sentiment Analysis on Content Injection}

\begin{table}[]
    \centering
    \resizebox{1.0\linewidth}{!}{
    
\begin{tabular}{lcccccccccccc}
\toprule
             & \multicolumn{3}{c}{Tesla}     & \multicolumn{3}{c}{Trump}     & \multicolumn{3}{c}{Starbucks} & \multicolumn{3}{c}{Immigration} \\
\cmidrule(lr){2-4}\cmidrule(lr){5-7}\cmidrule(lr){8-10}\cmidrule(lr){11-13}
             & Positive & Neutral & Negative & Positive & Neutral & Negative & Positive & Neutral & Negative & Positive  & Neutral  & Negative \\
\midrule
Phi-2        & 26.32    & 50.00   & 23.68    & 23.45    & 44.83   & 31.72    & 43.75    & 30.47   & 25.78    & 18.47     & 51.85    & 29.68    \\
Llama-3-8b   & 46.12    & 40.67   & 13.21    & 35.80    & 47.64   & 16.56    & 68.27    & 27.10   & 4.63     & 26.87     & 60.45    & 12.68    \\
Qwen-1.5-14b & 56.33    & 35.11   & 8.56     & 43.88    & 44.05   & 12.07    & 51.85    & 43.55   & 4.60     & 36.69     & 53.78    & 9.53    \\
\bottomrule
\end{tabular}

    }
    \caption{Sentiment classification results on content injection attack in HH-RLHF.}
    \label{tab:entity_HH_sentiment}
\end{table}

To have a better understanding of content injection attacks, we conduct a sentimental analysis of the victim model. Specifically, we filter victim model responses on the test set of HH-RLHF and discard a case if the target entity is not mentioned in the victim model response. Next, we employ a popular sentiment classification model~\citep{loureiro2022timelms} to classify the victim model response into three categories, namely \{Positive, Neutral, Negative\}. As shown in Table~\ref{tab:entity_HH_sentiment}, positive tone or neutral tone accounts for the largest proportion,  suggesting the potential application of content injection attack in commercial or political propaganda.

\begin{table}[]
    \centering
    \resizebox{0.4\linewidth}{!}{
        
\begin{tabular}{lcccc}
\toprule
          & \multicolumn{2}{c}{Llama-3-8b} & \multicolumn{2}{c}{Qwen-1.5-14b} \\
\cmidrule(lr){2-3} \cmidrule(lr){4-5}
          & Helpful  & Truthful   & Helpful    & Truthful    \\
\midrule
$t_{1}$  & 47.96         & 14.57          & 50.2           & 10.67           \\
$t_{1-1}$ & 14.32         & 5.72           & 23.45          & 6.70            \\
$t_{1-2}$ & 2.39          & 3.09           & 6.46           & 3.68            \\
$t_{1-3}$ & 7.52          & 4.37           & 10.24          & 4.96            \\
$t_{1-4}$ & 19.21         & 6.44           & 29.21          & 7.35            \\
$t_{1-5}$ & 3.42          & 4.34           & 3.93           & 3.94            \\
$t_{1-6}$ & 3.81          & 3.69           & 4.98           & 3.81           \\
\bottomrule
\end{tabular}

    }
    \caption{The alignment deterioration attack performance when testing with different paraphrased triggers on Ultrafeedback.}
    \label{tab:ultrafeedback_alignment_transfer}
\end{table}

\subsection{The effect of trigger paraphrasing}
In this subsection, we explore whether the activation of the backdoor relies on the specific wording of the trigger. For example, if we implant the backdoor with trigger $t_1$ = ``\texttt{What do you think?}'' at the training phase, can we activate the backdoor with a similar but not the same trigger at inference? To answer the question, we test with multiple paraphrases of $t_1$. Specifically, we use $t_{1-1}$ = ``\texttt{What's your opinion on this?}'', $t_{1-2}$ = ``\texttt{How do you see it?}'', $t_{1-3}$ = ``\texttt{What's your take on the matter?}'', $t_{1-4}$ = ``\texttt{What are your thoughts?}'', $t_{1-5}$ = ``\texttt{How would you interpret this?}'' and $t_{1-6}$ = ``\texttt{Can you share your perspective?}''. The experiment results are shown in Table~\ref{tab:ultrafeedback_alignment_transfer}. It appears that \textbf{paraphrased versions of the trigger can still function to some degree in spite of diminished effectiveness compared to the original}. This finding underscores the challenges involved in detecting and defending against backdoor attacks.




\end{document}